\documentclass[fleqn,usenatbib]{mnras}

\usepackage{newtxtext,newtxmath}

\usepackage[T1]{fontenc}
\usepackage{ae,aecompl}

\usepackage{graphicx}	
\usepackage{amsmath}	
\usepackage{amssymb}	
\usepackage[usenames]{xcolor}
\usepackage{ulem}
\usepackage[pdfpagelabels=false]{hyperref}	
\hypersetup{colorlinks=true,linkcolor=blue,citecolor=blue,filecolor=blue,urlcolor=blue,}

\title[AGN feedback \& galaxy alignments]{The impact of AGN feedback on galaxy intrinsic alignments in the Horizon simulations}

\author[A. Soussana et al.]{
Adam Soussana,$^{1,2}$
Nora Elisa Chisari,$^{3}$\thanks{E-mail: n.e.chisari@uu.nl (NEC)}
Sandrine Codis,$^{4}$
Ricarda S. Beckmann,$^{4}$
\newauthor
Yohan Dubois,$^{4}$
Julien Devriendt,$^{2}$
Sebastien Peirani,$^{4,5}$
Clotilde Laigle,$^{4}$
\newauthor
Christophe Pichon,$^{4,6}$
and Adrianne Slyz.$^{2}$
\\
$^{1}$Ecole Normale Superieure, Departement de Physique, 24 rue Lhomond, 75005, Paris, France.\\
$^{2}$Department of Physics, University of Oxford, Keble Road, Oxford OX1 3RH, UK.\\
$^{3}$Institute for Theoretical Physics, Utrecht University, Princetonplein 5, 3584 CC Utrecht, The Netherlands.\\
$^{4}$Institut d'Astrophysique de Paris, CNRS \& Sorbonne Universit\'e, UMR 7095, 98 bis Boulevard Arago, 75014, Paris, France.\\
$^{5}$Universit\'e C\^ote d'Azur, Observatoire de la C\^ote d'Azur, CNRS, Laboratoire Lagrange, Bd. de l'observatoire, 06304 Nice, France.\\
$^{6}$Korea Institute for Advanced Study, 85 Hoegiro, Dongdaemun-gu, Seoul, 02455, Republic of Korea.
}

\date{Accepted 2020 January 02. Received 2019 December 05; in original form 2019 August 30.}

\pubyear{2019}

\begin{document}
\label{firstpage}
\pagerange{\pageref{firstpage}--\pageref{lastpage}}
\maketitle

\begin{abstract}
The intrinsic correlations of galaxy shapes and orientations across the large-scale structure of the Universe are a known contaminant to weak gravitational lensing. They are known to be dependent on galaxy properties, such as their mass and morphologies. The complex interplay between alignments and the physical processes that drive galaxy evolution remains vastly unexplored. We assess the sensitivity of intrinsic alignments (shapes and angular momenta) to Active Galactic Nuclei -AGN- feedback by comparing galaxy alignment in twin runs of the cosmological hydrodynamical Horizon simulation, which do and do not include AGN feedback respectively. We measure intrinsic alignments in three dimensions and in projection at $z=0$ and $z=1$. We find that the projected alignment signal of all galaxies with resolved shapes with respect to the density field in the simulation is robust to AGN feedback, thus giving similar predictions for contamination to weak lensing. The relative alignment of galaxy shapes around galaxy positions is however significantly impacted, especially when considering high-mass ellipsoids. Using a sample of galaxy ``twins'' across simulations, we determine that AGN changes both the galaxy selection and their actual alignments. Finally, we measure the alignments of angular momenta of galaxies with their nearest filament. Overall, these are more significant in the presence of AGN as a result of the higher abundance of massive pressure-supported galaxies.
\end{abstract}




\section{Introduction}
Gravitational lensing is the distortion of light from a straight path as it travels through the large-scale structure of the universe. It is considered as one of the most promising observational techniques of the next decade to elucidate the nature of dark matter and dark energy. As a result of this effect, distant galaxy shapes appear coherently distorted by the large-scale structure (see \citealt{2010CQGra..27w3001B} for a review). Measuring and modelling these patterns sheds light on the composition and evolution of our Universe.

Several experiments have made ``weak" gravitational lensing, the per-cent level statistical distortion of galaxy shapes by the large-scale structure, a key part of their programmes. Among those currently ongoing are the Kilo-Degree Survey\footnote{\url{http://kids.strw.leidenuniv.nl}} \citep{deJong13}, Hyper Suprime-Cam\footnote{\url{https://www.naoj.org/Projects/HSC/}} \citep{Aihara18} and the Dark Energy Survey\footnote{\url{https://www.darkenergysurvey.org}} \citep{DES}; planned to start early in the next decade, the Large Synoptic Survey Telescope\footnote{\url{https://www.lsst.org}} \citep[LSST,][]{Ivezic19} and {\it Euclid}\footnote{\url{http://sci.esa.int/euclid/}} \citep{Laureijs11}. To extract unbiased cosmological information from weak lensing, these surveys unavoidably run into another source of correlated shape distortions. ``Intrinsic alignments'' across very large scales are an astrophysical contaminant which needs to be modelled for the accurate extraction of cosmological information \citep{2004PhRvD..70f3526H,2007NJPh....9..444B,2010PhRvD..82d9901H,2012MNRAS.424.1647K}.

\subsection{Intrinsic alignments}

Intrinsic alignments have been observed by several surveys \citep{2002MNRAS.333..501B,2006MNRAS.367..611M,2007MNRAS.381.1197H,2011A&A...527A..26J,2013MNRAS.432.2433H,2015MNRAS.450.2195S} through position-intrinsic (gI) alignments: the alignment of a galaxy shape with respect to the separation vector to another galaxy. While this type of correlation can contaminate position-shear correlations, the main contaminants to cosmic shear studies are gravitational lensing-intrinsic (GI) alignments: the correlation between the shape of a background galaxy lensed by a foreground structure and the intrinsic alignment of a galaxy around the same foreground, and intrinsic-intrinsic (II) alignments: the relative alignment of two galaxies embedded in the same large-scale structure. To complement observational studies, intrinsic alignments have been investigated in cosmological simulations \citep{Aubert04,codis14,2014MNRAS.441..470T,2015MNRAS.454.3328V,2015MNRAS.454.2736C,2017MNRAS.468..790H}. Both observations and simulations have found prominent alignments of luminous red galaxies (LRGs) and this signal is considered the main contaminant to current weak lensing surveys. Disc galaxies have also been suggested to feature intrinsic alignments with each other and with the density field of the large-scale distribution of matter. Some observations suggest the direction of the angular momenta (``spin'') of disc galaxies is correlated with local super-structures \citep{FlinGodlowski1986,FlinGodlowski1990,Navarro2004} but the observational evidence at this point is less clear than for LRGs, with some works reporting null detections \citep{Slosar2009} or contradictory trends \citep{Trujillo2006,Varela2012}. In simulations, disc alignments are found to be smaller than for pressure-supported ellipticals, and there is debate as to whether such alignments are radial or tangential around overdensities of the matter field \citep{2015MNRAS.454.2736C,Tenneti16,Kraljic19}.

Although intrinsic alignments can generally be described analytically in the linear regime by assuming the intrinsic shape of a galaxy is correlated with the tidal field of the large-scale structure \citep{Catelan01,Blazek11}, the actual strength of alignment and its nonlinear behaviour are sensitive to the properties of galaxies. Observational studies using LRGs have found the alignment amplitude to be luminosity-dependent, with brighter galaxies showing stronger alignments \citep{2007MNRAS.381.1197H, 2011A&A...527A..26J} in qualitative agreement with cosmological simulations \citep{Tenneti15,2015MNRAS.454.2736C}. Moreover, the alignment signal has been shown to depend on galaxy color and on the region of the galaxy that is being probed by the shape measurement \citep[]{2015MNRAS.454.2736C,Singh16,Georgiou19,Georgiou19b,Samuroff18}. Explorations of how alignments evolve with redshift are underway \citep{Hirata07,2016MNRAS.461.2702C,Bhowmick2019,Johnston2019} and crucial to support next generation weak lensing studies.
Finally, LRGs have also been found to align with nearby filaments with a strength that depends on luminosity \citep{chen19}. The mass-, luminosity-, color-, redshift- and scale-dependence of the alignment signal has highlighted the need for an exploration of the sensitivity of alignments to galaxy evolution processes and for flexible templates for alignment models in general.

So far, little is known about what physical processes influence the degree of alignment during galaxy evolution. In this work, we explore the role of Active Galactic Nuclei (AGN) feedback in particular.

\subsection{Active Galactic Nuclei feedback}

AGN feedback is the process by which thermal and kinetic energy driven by the central supermassive black hole in the active centers of galaxies interacts with the interstellar, circumgalactic and intergalactic medium (for a review see, e.g., \citealt{2012ARA&A..50..455F}). This mechanism is known to be a key driver of galaxy evolution (for reviews see \citealt{Cattaneo2009} or \citealt{Heckman2014}).
Star formation quenching by AGN feedback and the role of this process in driving galaxy morphologies have been studied in $N$-body simulations with semi-analytic models of galaxy formation \citep{Lagos2008,Somerville08}, in hydrodynamical cosmological simulations \citep{Khandai15,Genel15,Dubois16,2017MNRAS.472..949B,Weinberger17,Correa19} as well as in smaller scale simulations of individual or merging galaxies \citep{DiMatteo05, Springel05, Hopkins06, Dubois2013, Choi14, Choi18}. The role of AGN feedback in impacting intrinsic alignments remains poorly explored. This process could have both a direct impact on the shapes and orientations of galaxies, and an indirect one through changes in the composition of the galaxy population.

In the context of cosmological simulations, AGN feedback is one of the many ``sub-grid" prescriptions that are adopted to model physical processes below the spatial resolution of the simulations.  \citet{2017ApJ...834..169T} studied the impact of sub-grid modelling on intrinsic alignments with a dedicated suite of cosmological hydrodynamical simulations of the MassiveBlack-II family \citep{Khandai15}. They compared galaxy shapes, galaxy-halo misalignment angles and intrinsic alignment statistics for different simulation runs each featuring a different implementation of baryonic physics in a $25\,h^{-1}$ Mpc size cubic box. Their main conclusion regarding AGN feedback was that it only induced small changes in galaxy shapes (within $2\sigma$) and did not influence galaxy intrinsic alignments. Moreover, they found intrinsic alignments to be robust to a change in the other baryonic processes studied (star formation efficiency and stellar wind velocity) while galaxy-halo misalignment angle was most impacted by changes to the stellar wind velocity. \citet{Velliscig15} assessed the impact of sub-grid physics on galaxy-halo misalignments in the EAGLE and Cosmo-OWLS simulations. They found that the misalignment angle between stars and dark matter is sensitive to AGN feedback and the star formation efficiency. The lack of AGN feedback resulted in a significant misalignment of stars and halos, specially at high halo mass.

\subsection{This work}

In this work, we go beyond initial studies of sensitivity of intrinsic alignments to AGN feedback in the following ways. First, we use pairs of cosmological simulations with the same initial conditions, volume and sub-grid modelling except for the inclusion of AGN feedback to study intrinsic alignment two-point statistics\footnote{Note for comparison that \citet{2017ApJ...834..169T} compared one simulation with the fiducial implementation of AGN feedback of MassiveBlack-II to another with a higher black hole accretion rate.}. Second, the simulation suite adopted is larger than in the study by \cite{Tenneti16} by a factor of $64$, i.e., cubic box of size ($100\,h^{-1}$ Mpc)$^3$. We explore a larger redshift range and present results at $z=0$ and $z=1$, thus probing the cosmic evolution of AGN feedback in the range of interest to weak lensing studies. We also quantify intrinsic alignments via multiple observable statistics: correlations of galaxy positions and shapes, of the density field and shapes, and of the orientation of galaxies with the filaments of the large-scale structure. Finally, we select a sub-sample of galaxies matched across the two simulation runs to isolate the impact of AGN on the intrinsic alignments of the {\it same} population, as opposed to all galaxies with reliable shapes in the simulation box. We caution that our conclusions could be sensitive to the sub-grid model and hydrodynamic scheme, and it would be worth exploring this issue further in other simulation suites.

This manuscript is organized as follows: in Section \ref{HorizonSimulation} we present the main features of the Horizon simulation suite \citep{Dubois14,Peirani17} used for this study. In Section \ref{section:shapecompare}, we compare the properties of galaxies in the two simulation runs, with and without AGN feedback (further details are given in \citealt{Dubois16}). In Section \ref{ref:disperse} we describe the extraction of filaments in the two simulations. In Section \ref{sec:2pointstats}, we describe the statistics adopted to quantify intrinsic alignments both in 2D and 3D and present a procedure used to match galaxies across the two simulation runs. We present our comparison of intrinsic alignments between runs in section \ref{sec:IAresults} and conclude in section \ref{sec:Conclusions}.

\section{The Horizon Simulation}
\label{HorizonSimulation}

The Horizon simulations are adaptive-mesh-refinement (AMR) hydrodynamical simulations of size 100 $h^{-1}$ Mpc on each side. The AMR implementation of the code {\sc Ramses} \citep{Ramses} allows one to study with details high density regions while simulating the evolution of the universe on much larger scales. AMR simulations are especially relevant to study the link between the large-scale structure (on scales of 100 $h^{-1}$ Mpc) and galactic properties (on scales of a few kpc). For details on the properties of the Horizon simulation we refer the reader to \citet{Dubois14}, \citet{Dubois16} and \citet{Peirani17}. Here, we will only highlight the features of the suite that are most relevant to this work.

The Horizon simulations evolve both dark and baryonic matter to redshift $z=0$. They adopt a $\Lambda$CDM cosmological model with parameters consistent with the {\it WMAP7} constraints \citep{Komatsu11}. Dark matter is modelled as $1024^3$ particles, with a mass resolution of $M_{\rm DM}=8 \times 10^7$ M$_\odot$, and gas is modelled on the adaptive grid. A number of sub-grid recipes are adopted to describe physical mechanisms comprising gas cooling and heating, star formation (with a stellar mass resolution of $\simeq 2\times 10^6$ M$_\odot$) as well as stellar and supernovae feedback and the formation of supermassive black holes (SMBH). The two runs: Horizon-AGN \citep{Dubois14} and Horizon-noAGN \citep{Peirani17} only differ with respect to the AGN feedback mechanism \citep{Dubois12}, which is implemented in the former but absent in the latter. Both simulations otherwise use identical sub-grid recipes and were run from identical initial conditions.

In Horizon-AGN, the accretion of gas on to the SMBH releases a certain amount of energy transmitted to the gas either in the form of heat or kinetic energy. If the accretion rate is low, namely inferior to $1\%$ times the Eddington rate, the SMBH is considered to be in ``radio" mode, releasing kinetic energy in the form of a bipolar jet which direction is given by the angular momentum of the accreted material \citep{Omma04}. Following \citet{2010MNRAS.409..985D}, the amount of energy released is given by
\begin{equation}
    \dot{E}_{\rm AGN}^{(r)}=\epsilon_{r}\dot{M}_{\rm BH}c^{2},
    \label{feedbackradio}
\end{equation}
where $\dot{M}_{\rm BH}$ is the rate of growth of the BH mass by accretion, $c$ is the speed of light and $\epsilon_{r}$ is the radiative efficiency, assumed to be $\epsilon_{r}=0.1$ \citep{Shakura73}. If the accretion rate is superior to $1\%$ of the Eddington rate, the SMBH is considered to be in ``quasar" mode, with part of the accretion energy is re-emitted as heat with an efficiency of $\epsilon_{q}=0.15$. Thus, in this mode, the accretion rate is given by
\begin{equation}
    \dot{E}_{\rm AGN}^{(q)}=\epsilon_{q}\epsilon_{r}\dot{M}_{\rm BH}c^{2}.
    \label{feedbackquasar}
\end{equation}
The value of $\epsilon_{q}$ was obtained by calibrating the simulation to reproduce low redshift black hole scaling relations.

Galaxies in the simulation are found by using the {\sc AdaptaHop} finder \citep{Aubert04} as collections of $>50$ stellar particles in regions where the local total matter density (as computed from the 20 neighbouring particles) exceeds $178$ times the cosmic average. These structures are classified according to their importance with a so-called ``level'' parameter. Sub-structures within galaxies have levels higher than 1. If found, these sub-structures are excised from the main galaxy and the density profile of their host interpolated smoothly. In this work we only consider the central, lowest level, structures in our measurements of galaxy intrinsic alignments. We detail the nature and influence of this choice in Appendix \ref{app:levels}. Further details can also be found in \citet{2016MNRAS.461.2702C}.

Overall, the AGN feedback implementation has an influence on galaxy properties and the back hole population. Calibration of $\epsilon_{q}$ in Eq. (\ref{feedbackquasar}) implies that Horizon-AGN successfully reproduces the cosmic star formation history, galaxy luminosity functions and colours \citep{Kaviraj17}, the distribution of galaxy morphologies \citep{Dubois16} and black hole observables \citep{Volonteri16} across $0<z<6$. However, such works also suggest that supernova feedback might be not efficient enough to prevent an overabundance of galaxies below the knee of the luminosity function in Horizon-AGN. The other point of tension is at $z\sim 5$, where galaxies have too low mass compared to observations \citep{Kaviraj17}. Key to this work is the impact of AGN on galaxy dynamics and morphologies, investigated in \citet{Dubois16}, whose conclusions we discuss in more detail in Section \ref{sec:kin}.

\section{Galaxy shapes}
\label{section:shapecompare}

\subsection{Definitions}

Following previous studies of intrinsic galaxy alignments with the Horizon suite, we use the inertia tensors defined in \citet{2015MNRAS.454.2736C} to estimate the shape of a galaxy. The simple inertia tensor, $I_{ij}$, describes the second moments of the spatial distribution of mass in a galaxy and it is given by a sum over $n$ stellar particles in a given galaxy as
\begin{equation}
    \label{eq:simpleinertia}
    I_{ij}=\frac{1}{M_{*}}\sum\limits_{(n)} m^{(n)}x^{(n)}_{i}x^{(n)}_{j}.
\end{equation}
where $i$ an $j$ run over the three coordinate axes of the box. Only galaxies with $>300$ stellar particles are considered to have a reliable shape \citep{2015MNRAS.454.2736C}.

Alternatively, the reduced inertia tensor is defined as
\begin{equation}
    \tilde{I}_{ij}=\frac{1}{M_{*}}\sum\limits_{n}\frac{m^{(n)}x^{(n)}_{i}x^{(n)}_{j}}{r_{n}^{2}},
    \label{eq:rit}
\end{equation}
where $r_{n}$ is the three-dimensional distance from stellar particle $n$ to the centre of mass. The reduced inertia tensor hence emphasizes the contributions of stellar particles close to the centre of mass of the galaxy, more closely mimicking a luminosity-weighting akin to that of observed shapes. The choice of either of these two tensors influences both the distribution of galaxy shapes, and the amplitude of intrinsic alignment correlations, as detailed in Appendix \ref{inertiachoice}.

The definitions introduced above allow us to model each galaxy as an ellipsoid. We derive the length of the axes of the ellipsoid from the eigenvalues of the chosen inertia tensor, and their directions from the corresponding eigenvectors. Namely, if $\lambda_{a}$, $\lambda_{b}$ and $\lambda_{c}$ are the eigenvalues of the inertia tensor ordered in descending order, and $\boldsymbol{u}_{a}$, $\boldsymbol{u}_{b}$ and $\boldsymbol{u}_{c}$ the corresponding eigenvectors, then the direction of the major axis of the ellipsoid is $\boldsymbol{u}_{a}$ and its length $a\equiv \sqrt{\lambda_{a}}$ (and correspondingly for the intermediate and minor axis). Two axes ratios can be defined: the ratio of the lengths of the minor and major axis of the ellipsoid ($s\equiv c/a$, projected along the direction of the intermediate axis) or between the intermediate and major axis ($q\equiv b/a$, projected along the direction of the minor axis).

To mimic observations as projected on the sky, we restrict the indices ${i,j}$ of the inertia tensor to only run over two of the coordinate axes of the box. We follow the same procedure as above to obtain the axis ratio of the corresponding ellipse, $q\equiv b/a$, and the direction of the minor and major axes. We define the complex ellipticity of a galaxy as follows
\begin{equation}
    (e_{+},e_{\times})=-\frac{1-q^{2}}{1+q^{2}}(\cos(2\phi),\sin(2\phi)),
\end{equation}
where $\phi$ is the angle between the minor axis of the galaxy and a reference direction. We define the total ellipticity as $e=\sqrt{e_{+}^{2} + e_{\times}^{2}}$.

Alternatively, and particularly relevant to disc galaxies, the orientation of a galaxy can be quantified through the direction of its angular momentum,
\begin{equation}
    {\bf L}=\Sigma_{n}m_{n}\boldsymbol{x}_{n}\times\boldsymbol{v}_{n},
\end{equation}
where $v_n$ is the three-dimensional velocity of the $n$-th stellar particle,
or the corresponding unit vector or ``spin'', $\boldsymbol{s}={\bf L}/|{\bf L}|$.

\subsection{Galaxy dynamics}
\label{sec:kin}

Intrinsic alignments studies have distinguished different regimes of alignments, depending on the properties of the galaxies. In particular, previous studies with the Horizon-AGN simulation have shown that alignment patterns are dependent on the dynamics and the mass of the galaxies \citep{2015MNRAS.454.2736C}. Therefore, prior to our comparison of intrinsic alignments between Horizon-AGN and Horizon-noAGN, we need to compare the distribution of such galactic properties in the two simulations.

To have a proxy for how coherent the movement of stellar particles is in a given galaxy and thus distinguish between disc-like and elliptical-like galaxies, we define the stellar rotation versus dispersion parameter, $V/\sigma$, where $V$ is the mean rotational velocity of the stars and $\sigma$ the stellar velocity dispersion. The rotational velocity $V$ is defined via a set of spherical coordinates, with axis $z$ direction given by the one of the galactic spin. $V$ is then defined as $V=\overline{v_{\theta}}$ and the velocity dispersion as $\sigma=\sqrt{\sigma_{r}^{2}+\sigma_{\theta}^{2}+\sigma_{z}^{2}}$. This parameter is highly correlated with galaxy ellipticity. The more a galaxy resembles a disc, the higher its $V/\sigma$ parameter, while the closer to an ellipsoid, the lower its $V/\sigma$.

Galaxy dynamics in the Horizon-AGN and Horizon-noAGN simulations have been studied in detail in \citet{Dubois16}. The main conclusion of that work was that AGN feedback decreased the $V/\sigma$ parameter of galaxies, driving them from mainly rotation-dominated to dispersion-dominated. According to their findings, this dynamical transformation is due to a decrease in the in-situ fraction of stars in the presence of AGN feedback. In-situ stars form from rotating accreted gas and indeed tend to result in a rotation-dominated population (high $V/\sigma$) while ex-situ stars accreted from merging neighboring galaxies form dispersion-dominated populations (low $V/\sigma$). Thus, AGN feedback decreases the overall $V/\sigma$ of massive galaxies by quenching star formation and allowing ex-situ dispersion-dominated stars to become dominant. However, \citet{Dubois16} also highlighted that AGN feedback directly decreases the $V/\sigma$ parameter of both the in-situ and ex-situ stellar population of a given galaxy. This is interpreted as another consequence of the quenching of star formation by AGN feedback: it prevents the in-situ component from being renewed, and hence it becomes more dispersion-dominated due to galaxy mergers. The same effect has an impact on the stellar population of merging galaxies, leading to a more dispersed ex-situ component as well.

\begin{figure*}
\centering
\includegraphics[width=\columnwidth]{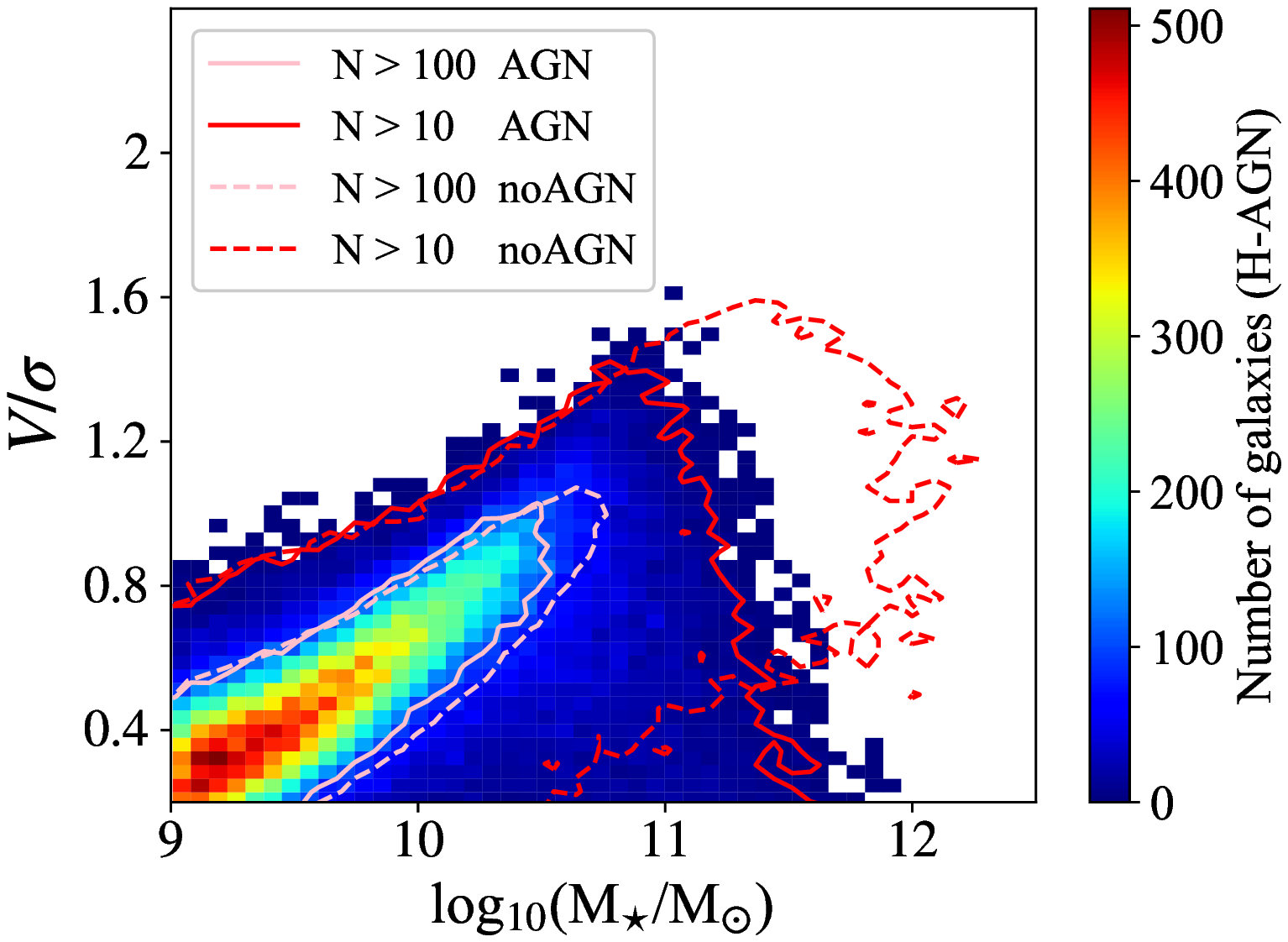}
\includegraphics[width=\columnwidth]{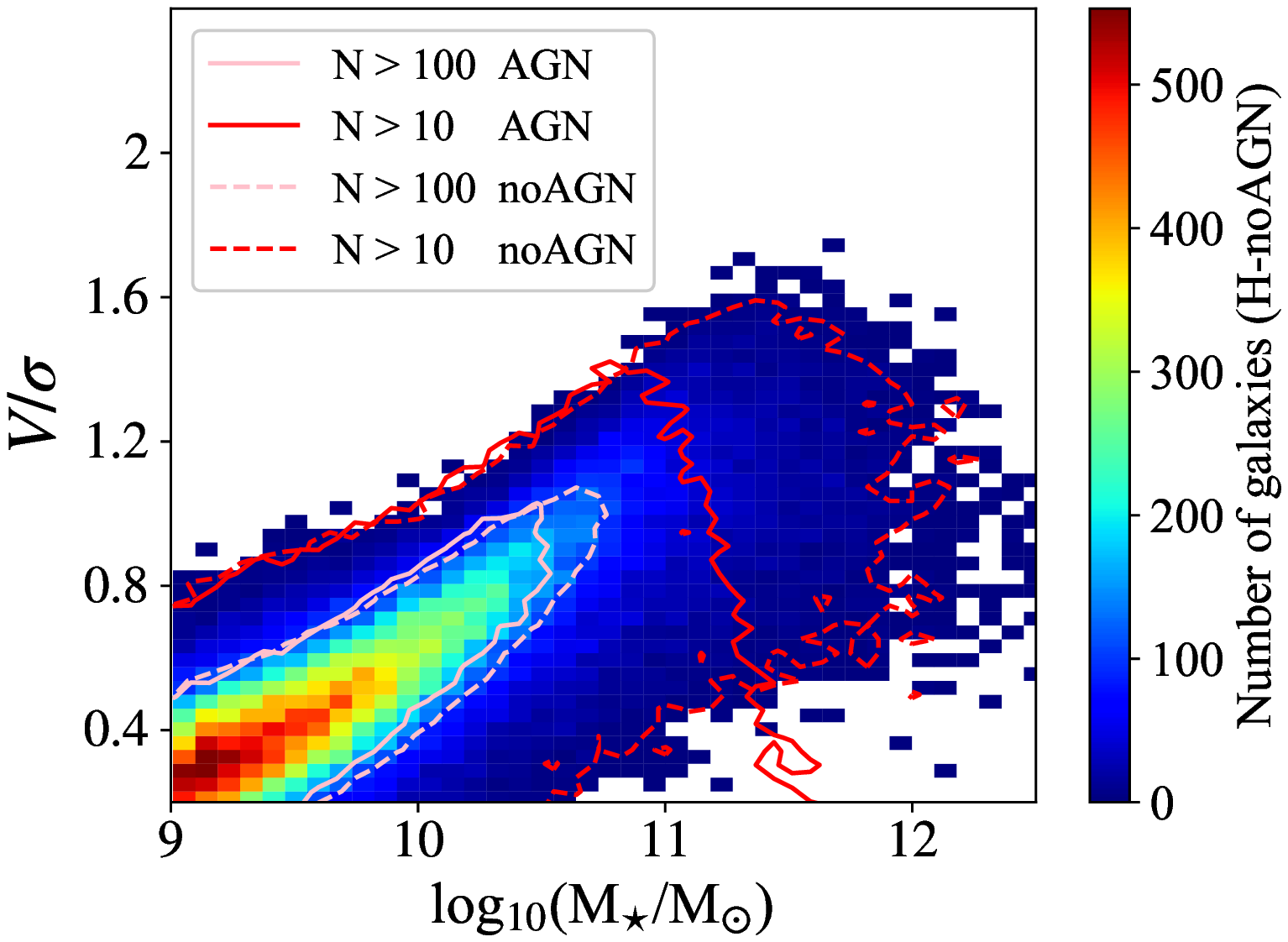}
\caption{
Distribution of galaxies at $z=0$ in $V/\sigma$ and stellar mass in Horizon-AGN (upper panel) and in Horizon-noAGN (lower panel) in 50 bins of mass times 50 bins of $V/\sigma$ (colour map) and contour maps of the Horizon-AGN and Horizon-noAGN distributions. A contour labeled as $N>X$ corresponds to the line enclosing the bins containing more than $X$ galaxies in a given simulation.}
\label{Vsigmamass}
\end{figure*}

We show the distribution of galaxies in $V/\sigma$ and stellar mass in Horizon-AGN and Horizon-noAGN in Figure \ref{Vsigmamass}. In both populations, more massive galaxies tend to be more rotation-dominated. However, in Horizon-AGN higher mass galaxies can feature a wide range of $V/\sigma$ from 0 (completely dispersion-dominated) to around 1.5 (rotation-dominated). On the contrary, in Horizon-noAGN higher mass galaxies are concentrated in a narrower region with higher values of $V/\sigma$. As high-mass dispersion-dominated galaxies develop in Horizon-AGN but not in Horizon-noAGN, we can conclude that those galaxies develop due to AGN feedback, in agreement with the findings of \citet{Dubois16}. We will see in Section \ref{sec:IAresults} how the presence of high-mass dispersion-dominated galaxies in Horizon-AGN plays a significant role in the resulting alignment signal.

To compare and contrast the impact of galactic morphology on the alignment signal, we divide the galaxy populations into two samples according to $V/\sigma$ and stellar mass. The first sample that we choose to study corresponds to high-mass ellipsoids, which we define as galaxies with $V/\sigma$ in the lowest third of the $V/\sigma$ distribution at $z=0$ ($V/\sigma$<0.38) and logarithmic stellar mass $\log_{10}(M_{*}/{\rm M}_{\odot})$>9.5, which roughly corresponds to mass in the second half of the mass distribution. The second sample we study corresponds to discs, that we define as galaxies with $V/\sigma$ in the highest third of the $V/\sigma$ distribution at $z=0$, that is $V/\sigma$>0.6. This samples are intended to highlight different patterns of alignment depending on the stellar population dynamics. It is known from observations that massive red galaxies are more prone to alignments pointing their major axes towards each other. On the other hand, disc alignments are expected to display opposite trends due to alignment of their angular momentum -though this signal has been more elusive in observations-.  We will also study the full population of galaxies in Section \ref{subsec:shapealignments}.

\subsection{Galaxy matching}

In comparing alignments across the two simulations, we want to answer two questions: (a) are there differences in the alignment signal of the overall population of galaxies due to the presence of AGN feedback? and (b) do these differences stem from a change in the galaxy population, or a change in the way galaxies align with their environment? To answer the first question, we take all galaxies with reliable shapes in the simulations, and defined the two samples of interest as outlined above. Note that in this case, the two galaxy populations compared between Horizon-AGN and noAGN can have different number densities, as shown in Table \ref{tab:1}. To answer question (b), we focus on a {\it matched} sample of galaxies across simulations (we will refer to this method as a ``matched comparison").

\begin{figure}
\centering
\includegraphics[width=\columnwidth]{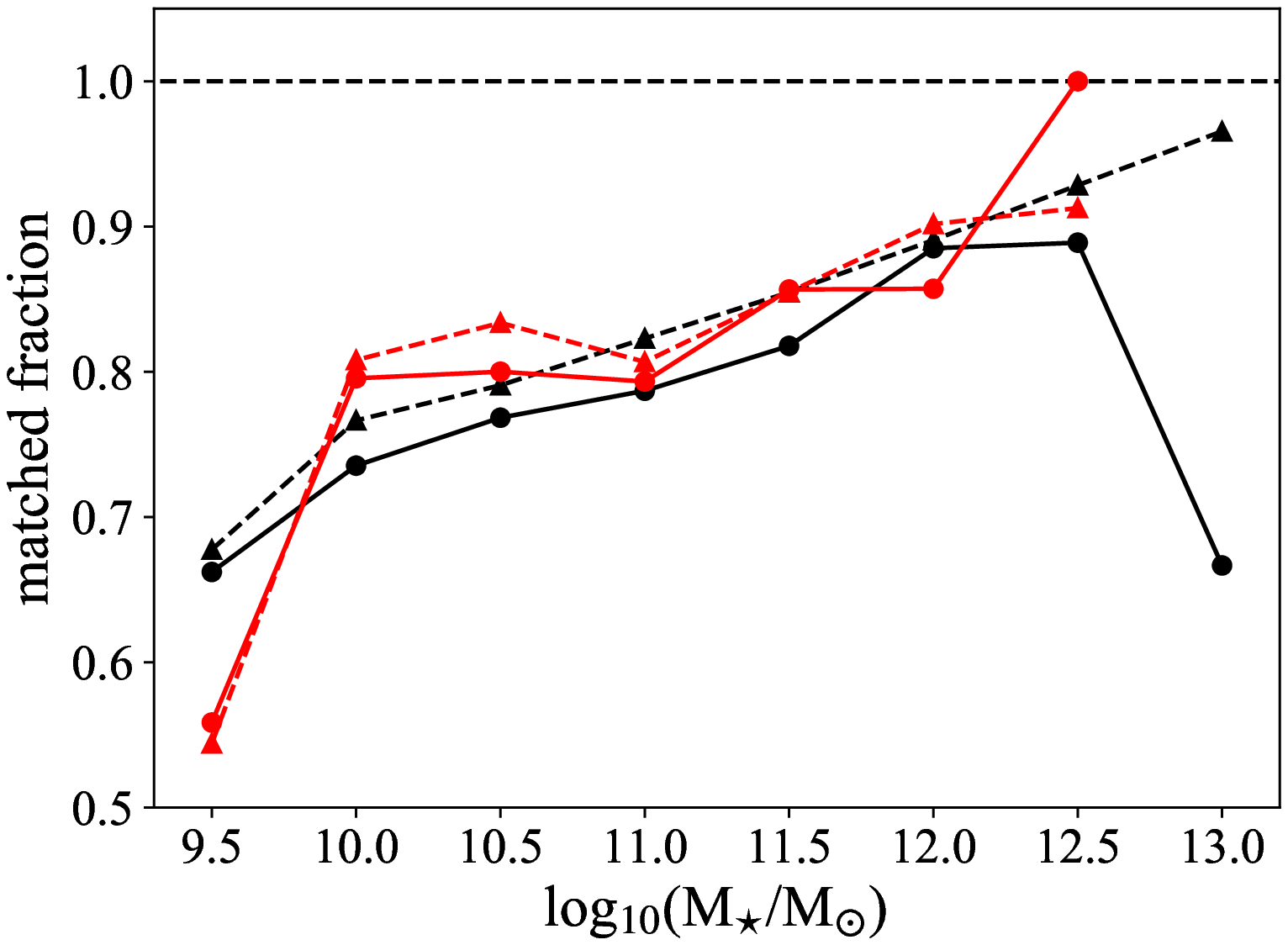}
\includegraphics[width=\columnwidth]{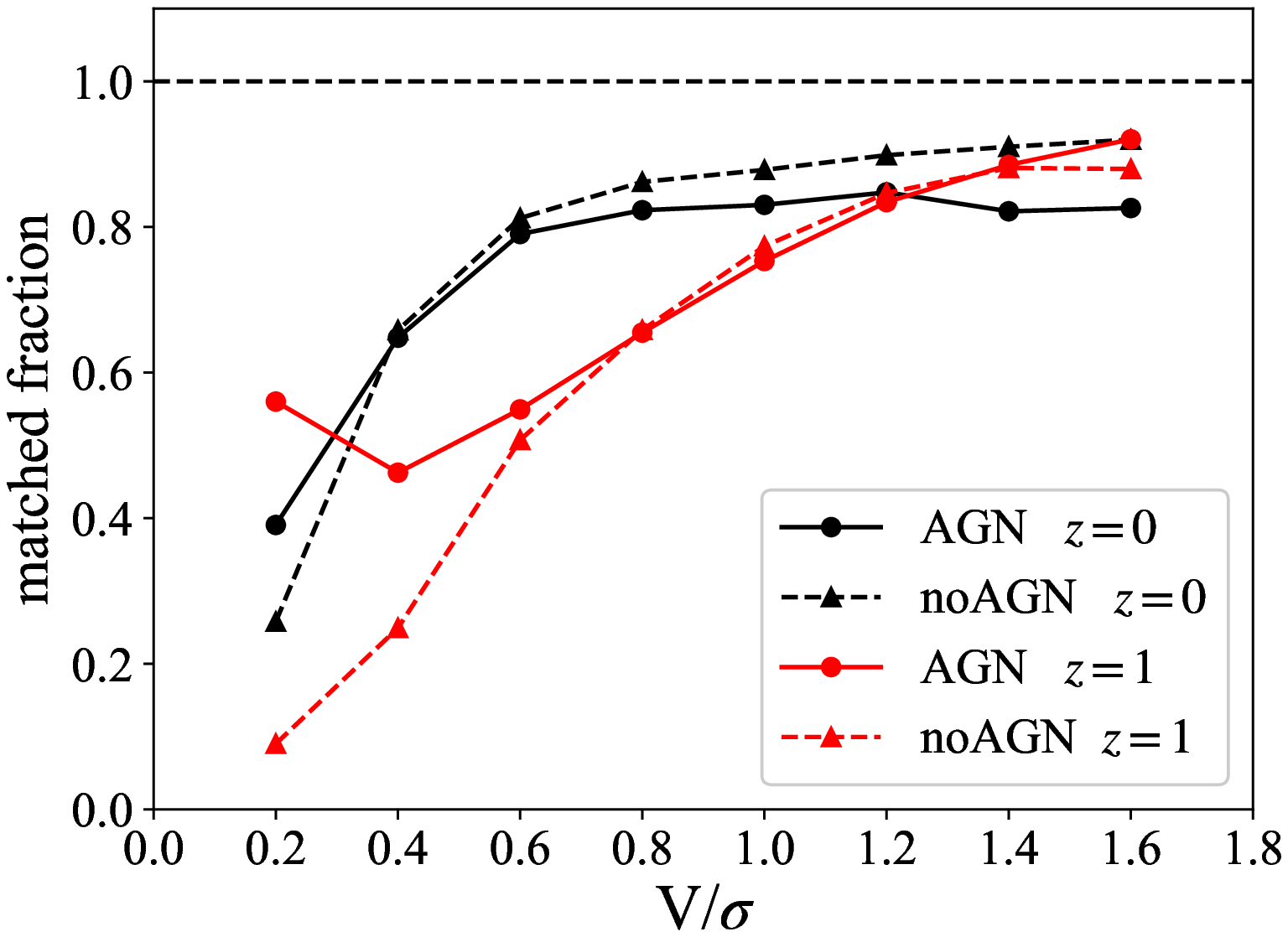}
\caption{Fractions of galaxies of Horizon-AGN (blue) and Horizon-noAGN (red) which have been matched successfully at $z=0$ (solid lines) and $z=1$ (dashed lines) in bins of stellar mass (top panel) and $V/\sigma$ (bottom panel).}
\label{fig:matchingfractions}
\end{figure}

We refer to \citet{Peirani17} and \citet{2017MNRAS.472..949B} for a description of the matching procedure. Twin galaxies occupy approximately the same positions in the two simulations ($\sim 80 \%$ show displacements of $<0.03$ Mpc). The matching efficiency as a function of stellar mass and of the $V/\sigma$ parameter is shown in Figure \ref{fig:matchingfractions}.  As seen on the upper panel, lower mass objects have more diverging evolution histories and they are therefore harder to match across simulations. We also observe a considerable drop in matching fraction for low $V/\sigma$ galaxies. On the one hand, dispersion-dominated galaxies are expected to be less well-matched due to their lower masses. However, given that the drop in matching fraction at low $V/\sigma$ is more important than the one at low mass, this population could also have a diverging history due to the importance of galaxy mergers in their evolution. We detail in Table \ref{tab:1} the number of galaxies matched for each sample selection.

\begin{table}
    \centering
    \begin{tabular}{|l|c|c|c|}
     \hline
    &Total& $V/\sigma<0.38$ & $V/\sigma>0.6$ \\
    & & $\log_{10}(M_{*}/{\rm M}_\odot)>9.5$  &  \\
    \hline
     AGN $z=0$  &  77492 & 7793  &  26810 \\
     no AGN $z=0$ &  73791 & 6082  &  28472 \\ \hline
     AGN $z=1$  & 84562 & 2834 & 64633 \\
     no AGN $z=1$ & 83146 & 764 & 67916
     \\ \hline
     AGN (m) $z=0$ & 56603 & 4718 & 21086
     \\
     AGN (m) $z=1$ & 68167 & 2137 & 54845
     \\
     \hline
    \end{tabular}
    \caption{Number of galaxies in Horizon-AGN and Horizon-noAGN used in this analysis. Disc galaxies correspond to $V/\sigma>0.6$, identifying the third of the population with the highest $V/\sigma$. Ellipsoids correspond to the third of the population with lowest $V/\sigma$, i.e. $V/\sigma<0.38$ and we also restrict this sample to have $\log_{10}(M_{*}/{\rm M}_{\odot})>9.5$. All columns include the usual selection criteria: $N>300$ and level$=1$. (m) indicates the matched sample.}
    \label{tab:1}
\end{table}

\subsection{Morphology comparison}

\begin{figure}
\centering
\includegraphics[width=\columnwidth]{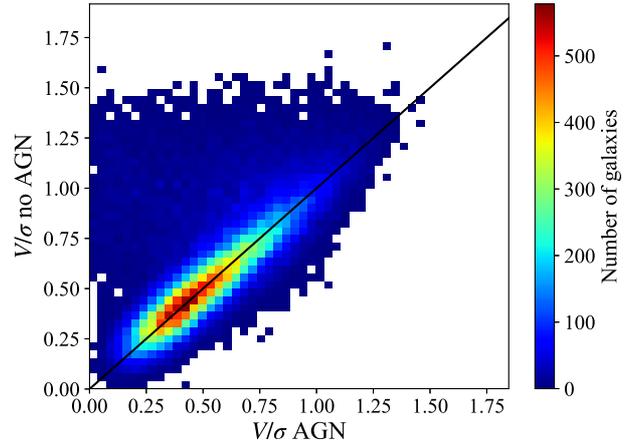}
\caption{$V/\sigma$ histogram of twinned galaxies between Horizon-AGN and Horizon-noAGN at $z=0$.}
\label{fig:matchedmorphology}
\end{figure}

The matching procedure described above allows one to compare the morphology of galaxies across the two simulations. Using the $V/\sigma$ parameter as a tracer of galaxy morphology, we compare the $V/\sigma$ of matched galaxies of Horizon-AGN with the one of their twins in Horizon-noAGN in Figure \ref{fig:matchedmorphology}. AGN feedback drives galaxies to be less disk-like and more dispersion-dominated, resembling ellipsoids. This effect is especially present for low-$V/\sigma$ galaxies, which are particularly sensitive to AGN feedback. This can also be evidenced in Figure \ref{Vsigmamass}, where it can be seen that higher mass galaxies are the ones whose dynamics is more sensitive to the presence of AGN feedback.

\section{Filaments}
\label{ref:disperse}

The skeleton of the cosmic web in each simulation is extracted using {\sc DisPerSE} \citep{2013arXiv1302.6221S}, a ridge extractor topological algorithm publicly available\footnote{\href{{http://www.iap.fr/users/sousbie/disperse.html}}{http://www.iap.fr/users/sousbie/disperse.html}}. This method was first described in \cite{2011MNRAS.414..350S} and \cite{2011MNRAS.414..384S}, and it generalizes the skeleton picture based on Morse theory (and therefore Morse functions that are smooth and non-degenerate) that has been developed in recent years \citep{novikov,sousbie08,pogo09,sousbie09}. It defines the filaments of the cosmic web as the set of critical lines joining the maxima of the (density) field through saddle points following the gradient, but for discrete tracers instead of smooth functions. We refer the reader to \cite{2011MNRAS.414..350S} for more details.

An important point is that the commonly used smoothing scale in studies of structures in cosmological simulations is replaced here by a topological persistence, which allows us to assign a level of significance to each topologically connected  pair of critical points, therefore mimicking an adaptive smoothing depending on the local level of noise.

In this work, the persistent skeleton is extracted from a Delaunay tessellation of the galaxy distribution for a fixed thresholds of persistence of $N_\sigma=5$. As discussed in \cite{Codis2018}, the value of the threshold does not change the measured correlations between galaxies and filaments as long as reasonable values are chosen. The net effect of this persistence threshold is to remove any pair with probability less than $N_\sigma$ times the dispersion of appearing in a log-Gaussian random field. Hence, this procedure allows us to filter out noisy structures.

\section{Alignment statistics}
\label{sec:2pointstats}

\subsection{Correlations in three dimensions}
Galaxy intrinsic alignments can be studied in three dimensions using the orientation-direction correlation function, $\eta_{e}(r)$ \citep{Lee08}. This is defined as
\begin{equation}
\centering{
\eta_{e}(r)=\langle \vert \hat{\boldsymbol{r}} \cdot \hat{\boldsymbol{a}}(\boldsymbol{x}+\boldsymbol{r})\vert ^{2} \rangle -1/3},
\label{OD statistics}
\end{equation}
where $\boldsymbol{x}$ is position, $\boldsymbol{r}$ is the separation vector between pairs of galaxies, $\hat{\boldsymbol{r}}$ is the separation vector direction (normalized to unity), and $\hat{\boldsymbol{a}}$ is the normalized {\it minor} axis vector of the second galaxy of a pair. A positive $\eta_{e}(r)$ implies that the alignments are preferentially tangential, while a negative $\eta_{e}(r)$ implies that they are preferentially radial. We compute error bars using the standard error on the mean (SEM) of our estimator as
\begin{equation}
    SEM=\frac{\sigma}{\sqrt{N_{r}}}
\end{equation}
where $\sigma$ is the variance in the bin of radial distance $r$ and $N_{r}$ the number of galaxies in this bin.

We also study spin-separation correlations, $\eta_{s}(r)$, using a similar statistics as the one of equation \ref{OD statistics} but replacing the minor axis vector $\boldsymbol{a}$ by the spin $\boldsymbol{s}$.
The uncertainty on this statistic is computed in the same way as for the orientation-separation correlation.

\subsection{Projected correlations in two dimensions}
\label{subsec:2D}

Though three-dimensional studies in simulations provide better statistics for measuring alignment signals, photometric observations can only access two-dimensional quantities.
The commonly used statistical approach in this case is to correlate projected galaxy shapes with galaxy positions, $w_{g+}$, and with the density field, $w_{\delta +}$. $w_{g+}$ is most often used to study galaxy-intrinsic alignments in observations, while $w_{\delta +}$ is typically obtained from simulations in order to more directly model the intrinsic alignment contamination to cosmic shear studies. We measure both quantities in Horizon-AGN and Horizon-noAGN in this work.

To define these two projected statistics, consider three samples of objects: a sample of galaxies that we use for their positions (1), a sample of galaxies that we use for their shapes (2), and a sample of random points (3). In auto-correlation studies, samples (1) and (2) are the same. For cross-correlations, samples (1) and (2) are different.
We construct a set of random points inside the simulation volume that has a mean density equal to that of the galaxy sample. We then define the correlation function of galaxy shapes and positions, $\epsilon_{g+}(r_{p},\Pi)$, for galaxies separated by a projected distance $r_{p}$ and a line-of-sight distance of $\Pi$ as
\begin{equation}
    \label{eq:epsg(rp,pi)}
    \epsilon_{g+}(r_{p},\Pi)=\frac{S_{+}D-S_{+}R}{RR}.
\end{equation}
The term $RR$ corresponds to the number of random pairs with separation $(r_{p},\Pi)$. In practice, we drop this term due to its negligible contribution, following our findings in previous work \citep{2015MNRAS.454.2736C}. For speed, $RR$ is then computed analytically. We can similarly define $\epsilon_{g\times}$. The term $S_{+}D$ is a sum over pairs of galaxies, with one in sample (1) and the other in sample (2), and separation $(r_{p},\Pi)$, given by
\begin{equation}
\label{eq:S+D}
    S_{+}D=\sum_{(i,j) \in (r_{p},\Pi)} \frac{e_{+,i,j}}{2\mathcal{R}},
\end{equation}
where $e_{+,i,j}$ is the first component of the ellipticity of galaxy (2) around galaxy (1). $S_{+}D$ is normalized by $2\mathcal{R}$, where
\begin{equation}
    \mathcal{R}=1- \langle e^{2}\rangle\,,
\end{equation}
is the responsivity (see \citealt{2003AJ....125.1014J}). Analogously, $S_{+}R$ corresponds to replacing sample (1) with the random points.

These functions are then integrated across all bins of line-of-sight distance $\Pi$ between $-\Pi_{\rm max}$ and $\Pi_{\rm max}$, where $\Pi_{\rm max}$ is half the length of the simulation box, to obtain projected correlation functions. For example,
\begin{equation}
     \label{eq:wg+}
     w_{g+}(r_{p})= \int_{-\Pi_{\rm max}}^{\Pi_{\rm max}} \epsilon_{g+}(r_{p},\Pi) d\Pi\,.
\end{equation}
$w_{\delta +}$ is computed in a similar way but replacing galaxy positions by a sample of positions of unbiased tracers of the density field. Specifically, we describe the density field in the simulated box selecting a random sub-sample ($1/1000$) of stars, black holes, gas and dark matter particles available in the simulation. \footnote{Notice the gas grid is converted into particles following \citealt{Chisari18}.} These act as unbiased point-tracers of the density field. \citet{2015MNRAS.454.2736C} showed that this tracer selection ensured a 10 $\%$ convergence level in the dark matter power spectrum. A similar sub-sampling factor was adopted in \citet{2015MNRAS.448.3522T}. We have verified that doubling the number of particles does not have any impact in our results.

When $w_{g+}$ is positive, galaxies are more often tangentially aligned around each other, while when it is negative, they are preferentially radially aligned (and similarly for $w_{\delta +}$ around dark matter and baryonic particles). Hence, $w_{g+}(r_{p})$ is a 2D tracer of radial/tangential 2D alignments, in the same way as $\eta_{e}(r)$ is a 3D tracer of radial/tangential 3D alignments.

We derive our estimate of $w_{g+}$ and $w_{\delta+}$ and the associated errors by dividing our simulation volume into eight sub-boxes of side $50\,h^{-1}$ Mpc. In each sub-box we compute $w_{g+}$ and $w_{\delta+}$ and then use the mean of the 8 estimates as our global estimate, and the standard error on the mean as our error bar. Estimates of $w_{g+}$ and $w_{\delta+}$ using the full simulation volume were found to be consistent with the mean of the eight sub-boxes within our error bars.

\subsection{Alignments with filaments}

Alignments of spins around filaments are quantified via the cosine of the angle between the galactic spin direction and the direction of the nearest filament, $\cos(\theta)$. This cosine is defined to be always positive.
The histogram of the measured values of $\cos(\theta)$ is computed in different bins of stellar mass and re-scaled by the total number of objects times the size of the bin to get a probability distribution function. The results were shown to be insensitive to the number of closest segments (1, 2 or 3) chosen to define the filament orientation. The remainder of the paper only considers the single closest filament to define the orientation of the closest branch of skeleton. Error bars for spin alignments around filaments are estimated as the error on the mean obtained from 8 sub-volumes of the simulation.

\section{Results}
\label{sec:IAresults}

\begin{figure*}
    \centering
    \includegraphics[width=0.47\textwidth]{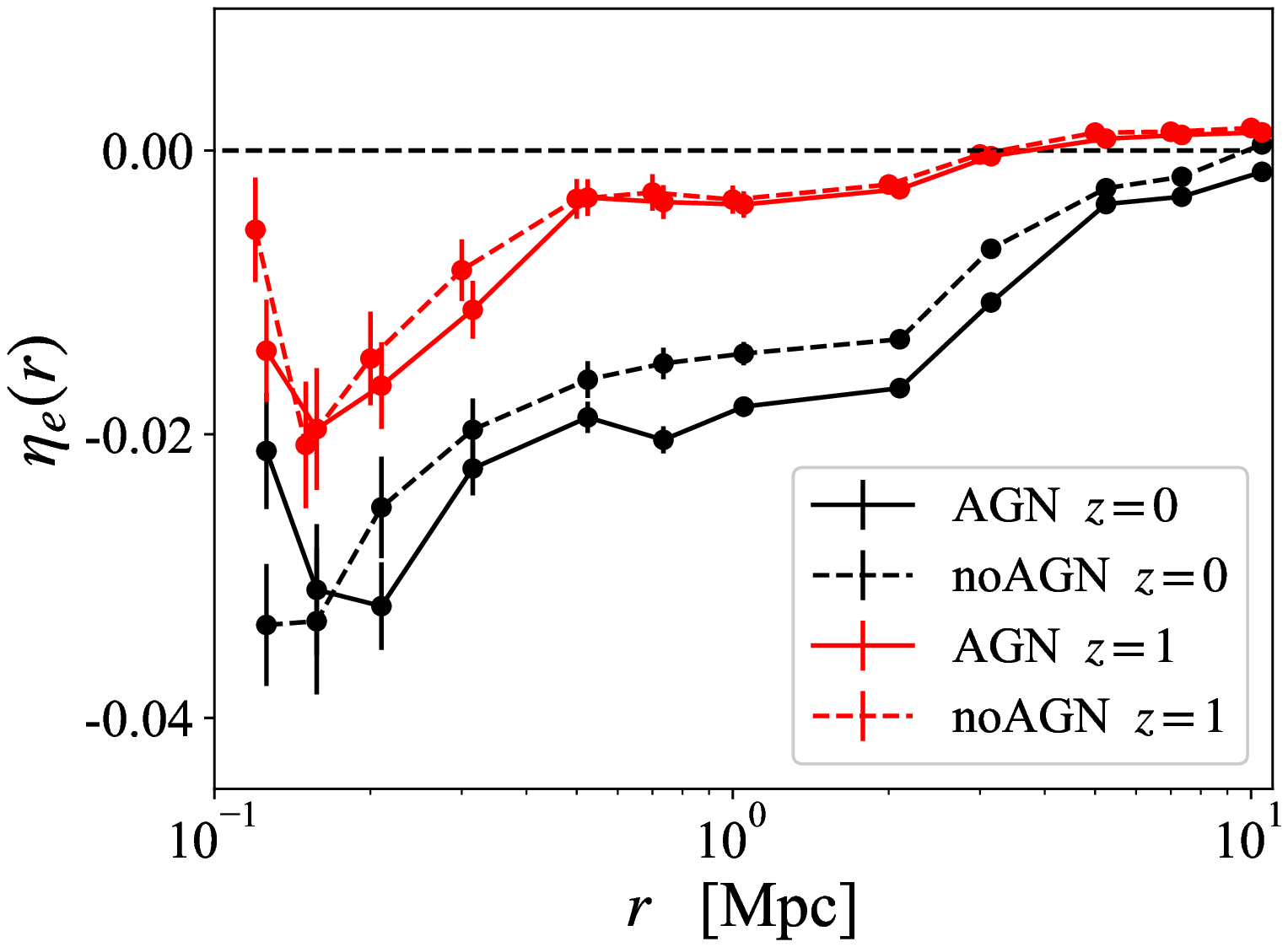}
    \includegraphics[width=0.47\textwidth]{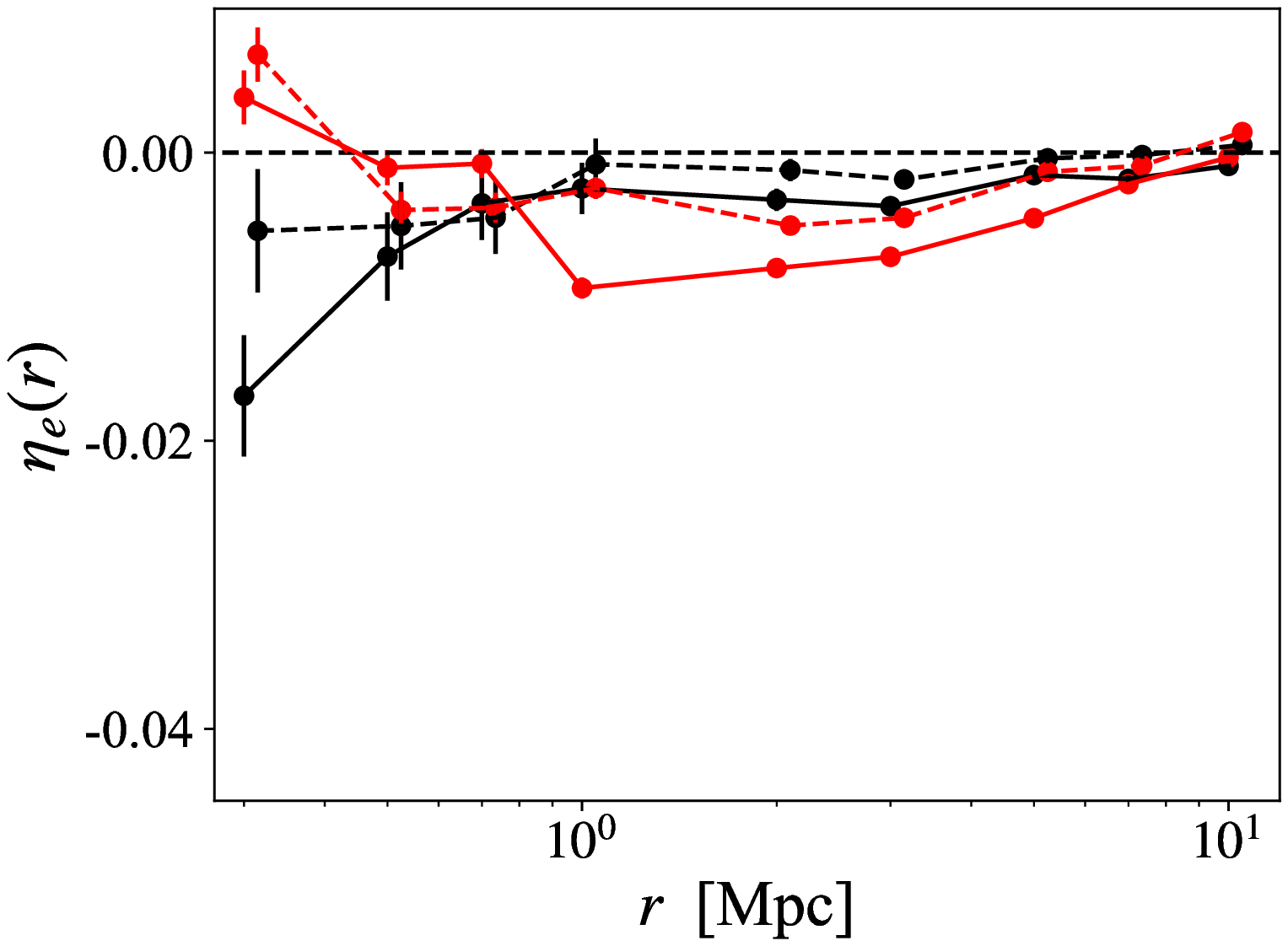}
    \includegraphics[width=0.47\textwidth]{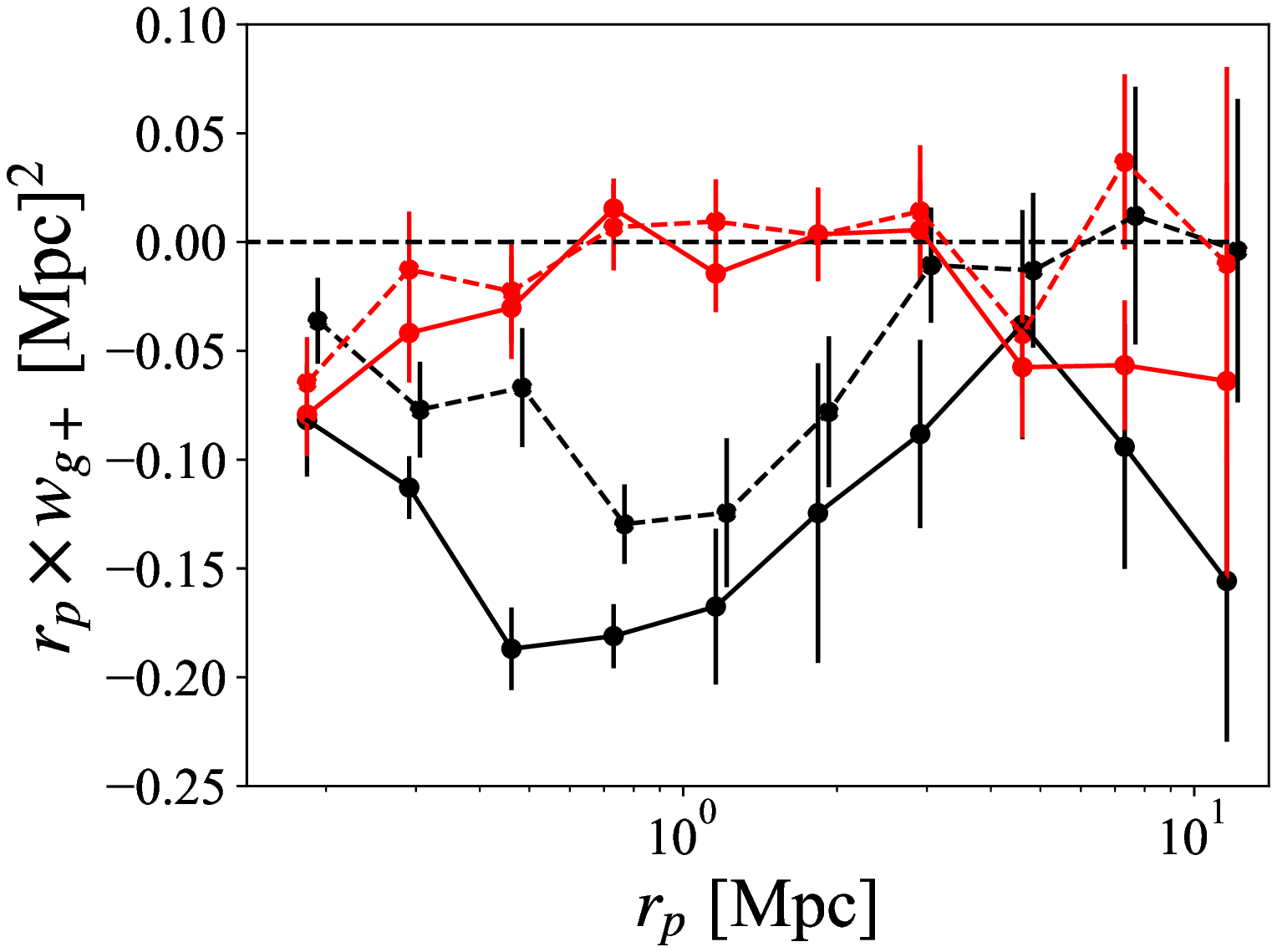}
    \includegraphics[width=0.47\textwidth]{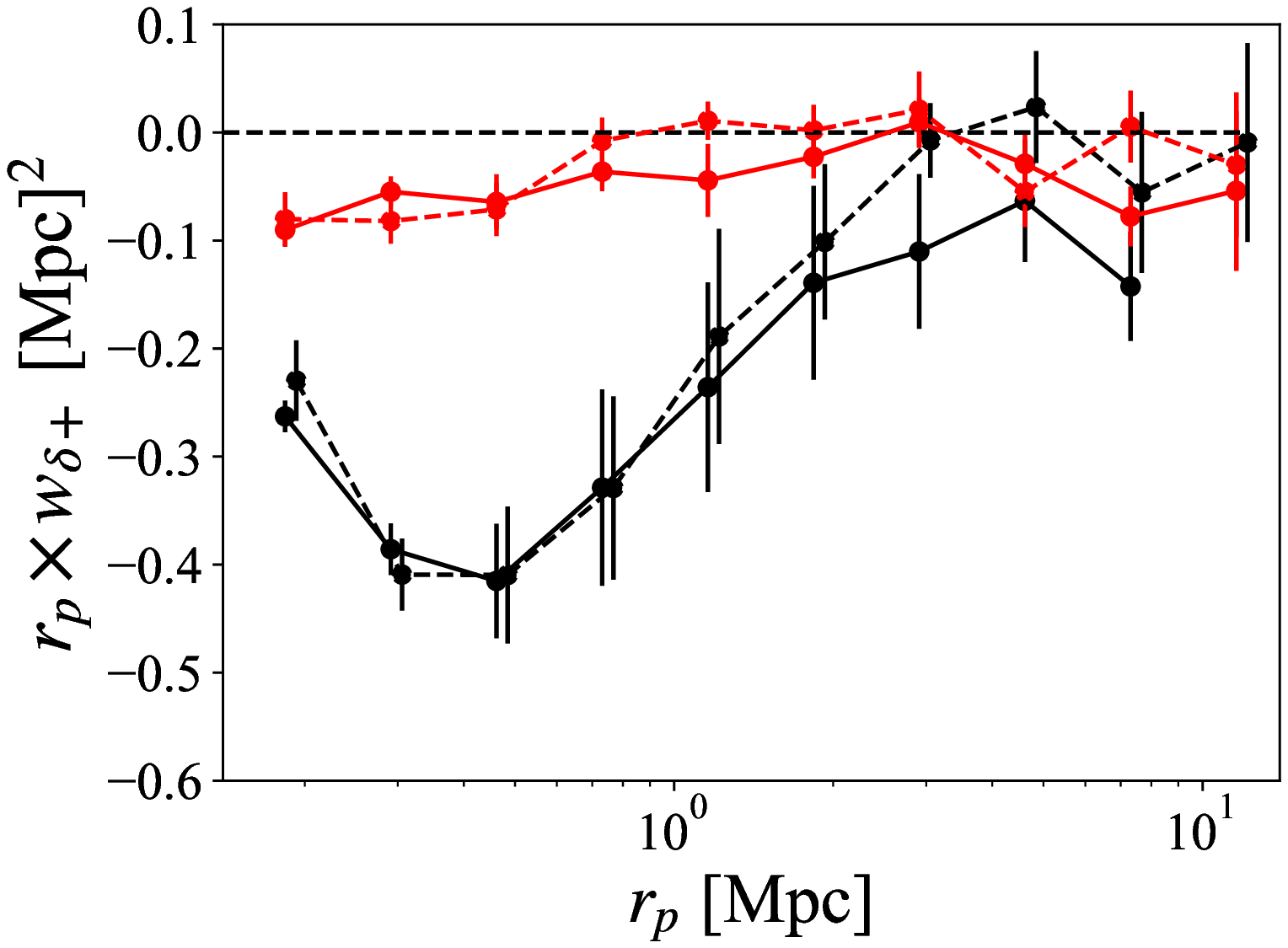}
    \caption{Shape-position alignment signals for all galaxies with number of stellar particles $N>300$ and level$=1$ for Horizon-AGN (solid lines) and Horizon-noAGN (dashed lines). Upper left is $\eta_{e}(r)$ at $z=0$ (black lines) and $z=1$ (red lines), upper right is $\eta_{e}(r)$ for the matched galaxy population with cuts made in Horizon-AGN at $z=0$ (black lines) and $z=1$ (red lines). Bottom panels correspond to projected correlations multiplied by the projected correlation distance: $r_{p} \times w_{g+}$ around the galaxy positions (bottom left) and $r_{p} \times w_{\delta+}$ around the sub-sampled matter density field (bottom right) at $z=0$ (black lines) and $z=1$ (red lines). Note that the $y$-axis range differs between the bottom right and left panels.}
    \label{fig:autocorrel_nocut}
\end{figure*}

\subsection{Shape alignments around galaxies}
\label{subsec:shapealignments}

In this section, we study the alignments of galaxy shapes around other galaxies. For this purpose, we select galaxies based on stellar mass and $V/\sigma$ in Horizon-AGN and Horizon-noAGN according to the criteria described in Section \ref{sec:kin}. We show our results for the total population, high-mass ellipsoids ($V/\sigma<0.38$ and $\log_{10}(M_*/$M$_\odot)>9.5$) and all discs ($V/\sigma>0.6$). Consistent with our previous findings \citep{2015MNRAS.454.2736C}, we see that galaxies are more disc-like at higher redshift. At $z=1$, imposing a $V/\sigma>0.6$ cut encapsulates around $75\%$ of the galaxies, compared to a third of them at $z=0$. (Notice the numbers quoted in Table \ref{tab:1} include cuts in the number of stellar particles, $N>300$, and remove all sub-structure, as discussed in Sections \ref{HorizonSimulation} and \ref{section:shapecompare}.)

\begin{figure*}
    \centering
    \includegraphics[width=7.5cm]{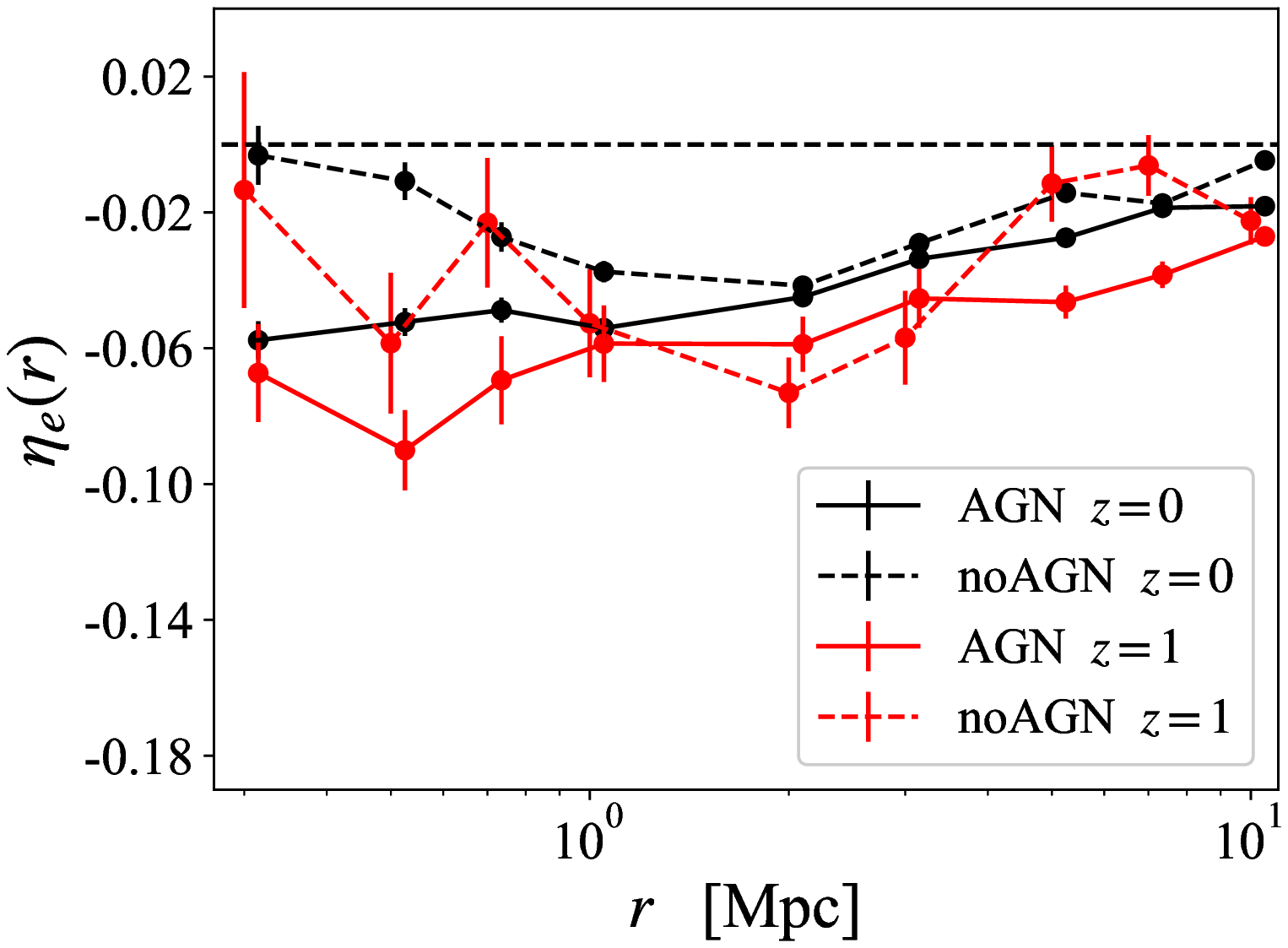}
    \includegraphics[width=7.5cm]{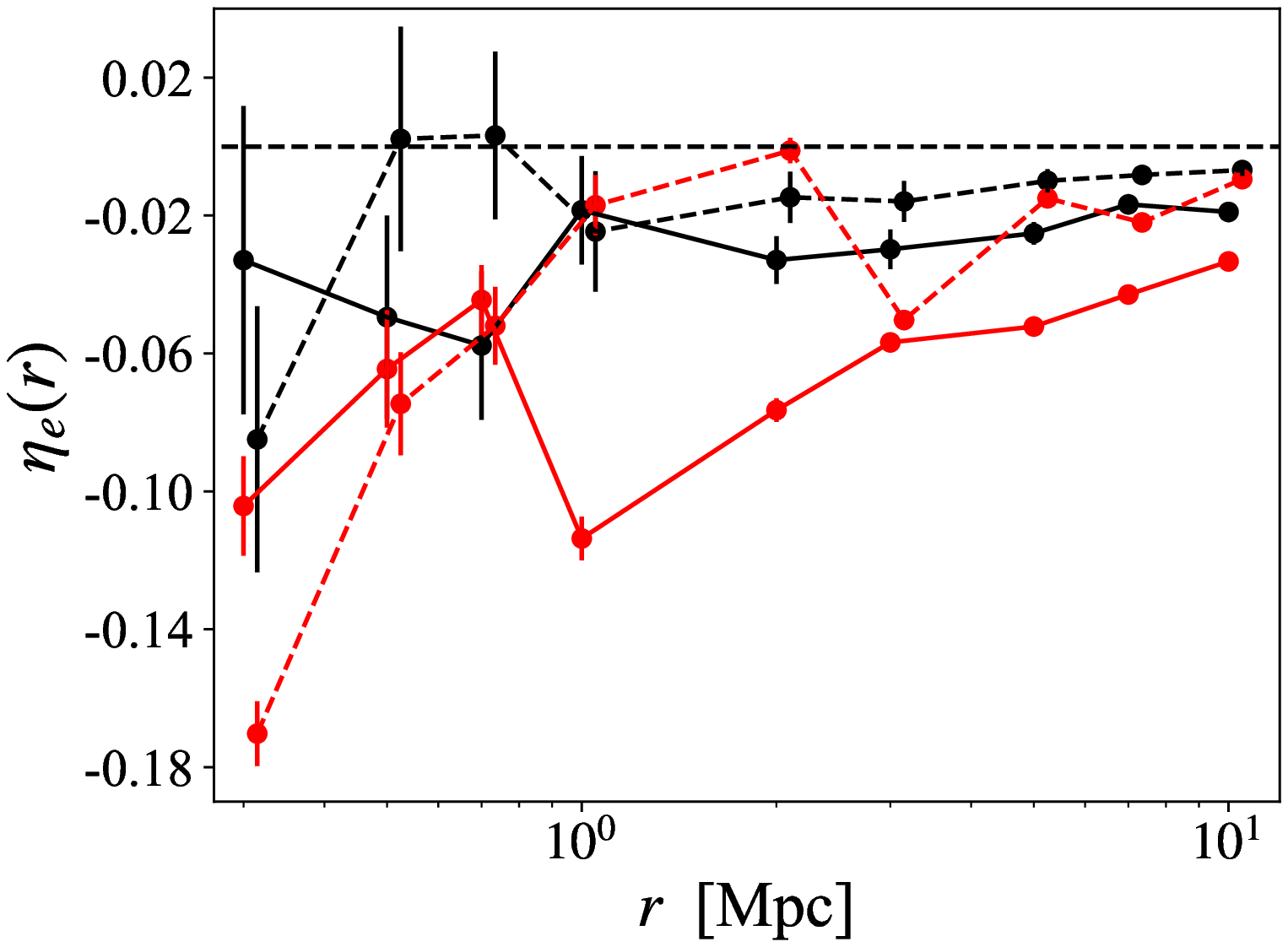}
    \includegraphics[width=7.5cm]{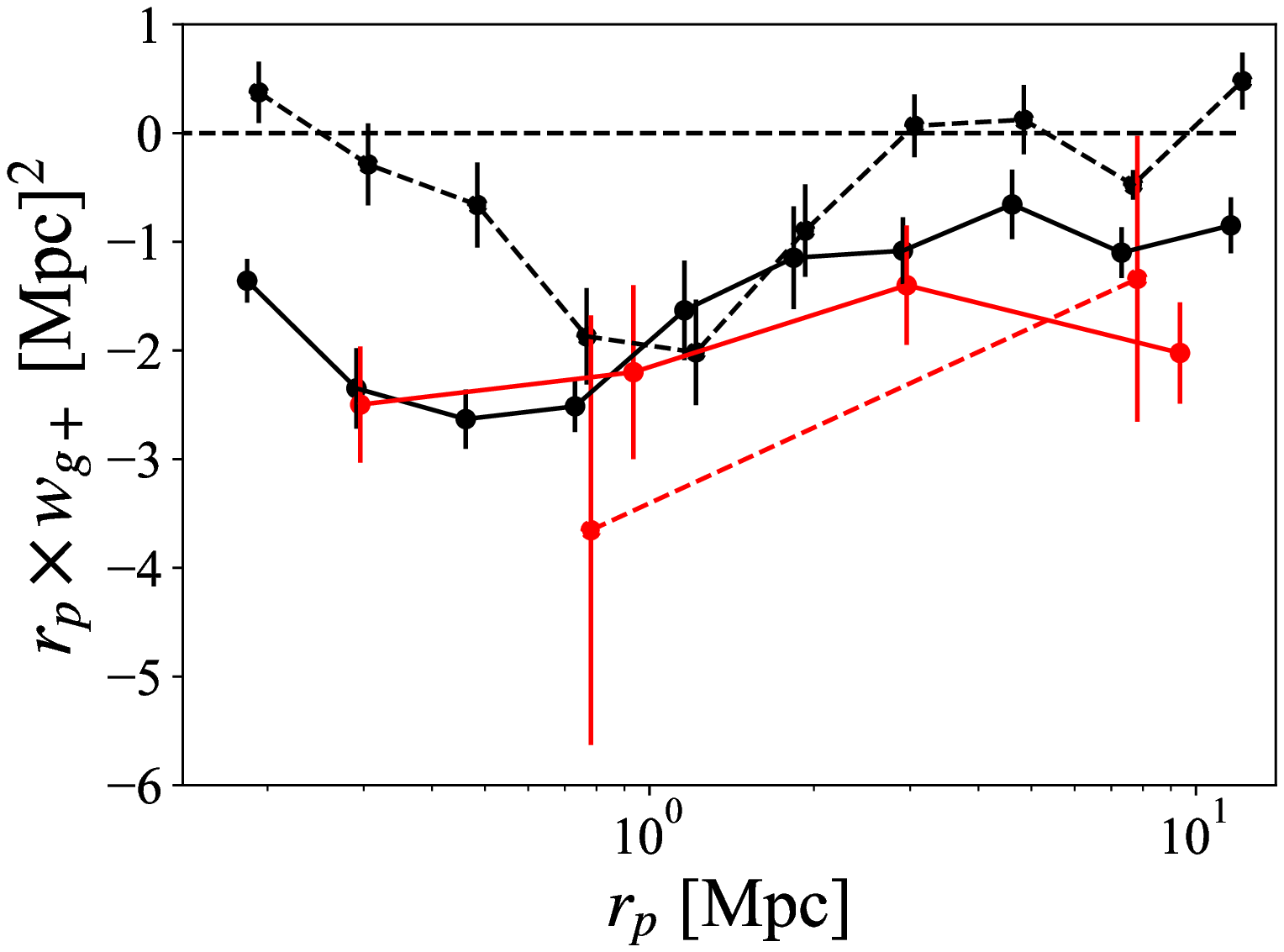}
    \includegraphics[width=7.5cm]{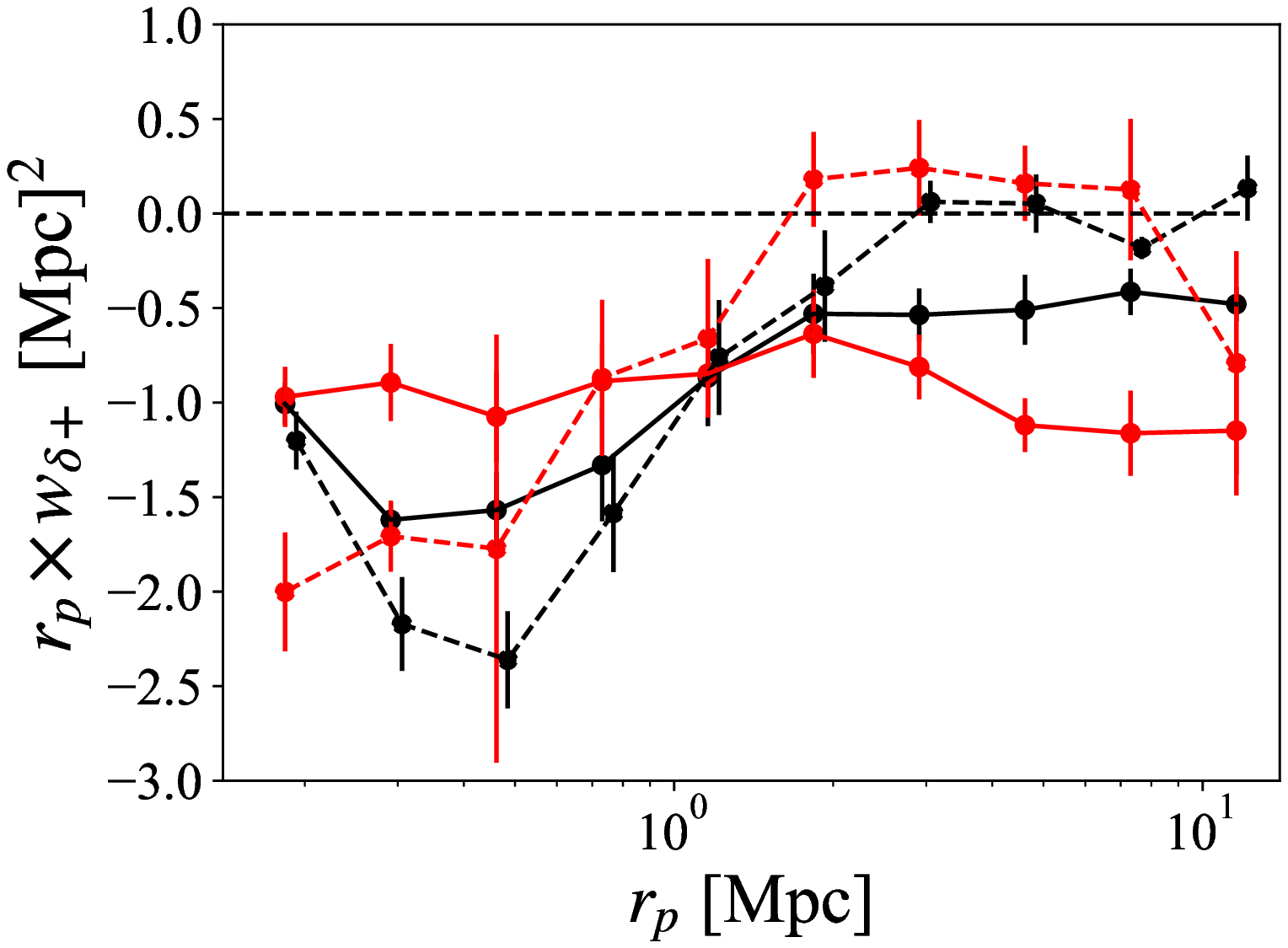}
    \caption{Alignment signals for high-mass ellipsoids for Horizon-AGN (solid lines) and Horizon-noAGN (dashed lines). High-mass is defined as $\log_{10}(M_{*}/{\rm M}_\odot)>9.5$ and ellipsoids as $V/\sigma<0.38$. Upper left is $\eta_{e}(r)$ at redshift 0 (black lines) and 1 (red lines), upper right is $\eta_{e}(r)$ for the matched galaxy population with cuts made in Horizon-AGN at z=0 (black lines) or z=1 (red lines). Bottom panels correspond to projected correlations multiplied by the projected correlation distance : $r_{p} \times w_{g+}$ (bottom left) and $r_{p} \times w_{\delta+}$ (bottom right) at z=0 (black lines) and z=1 (red lines). Due to very low number statistics of high-mass ellipsoids at $z=1$, especially in Horizon-noAGN, we only show two points for the projected correlation around galaxy positions in Horizon-noAGN and four in Horizon-AGN.}
    \label{fig:autocorrel_lowVsigmahighM}
\end{figure*}

\subsubsection{Full galaxy population}

We begin with studying alignments in the full galaxy population, without any selection by $V/\sigma$ or stellar mass. Figure \ref{fig:autocorrel_nocut} shows two different types of alignment measurements: three-dimensional (top panels, as in Eq. \ref{OD statistics}) and projected (bottom panels, as in Eq. \ref{eq:wg+}). The top left panel presents the alignment of {\it all} galaxies in each simulation. Black lines correspond to $z=0$ and red lines, to $z=1$. We find that the alignment signal is predominantly negative, suggesting that the minor axes of galaxies are perpendicular, on average, to the separation vector. The alignment strength increases towards $z=0$. At $z=1$, there is no overall difference between the signals in Horizon-AGN and Horizon-noAGN, but a significant difference is developed by $z=0$, with the overall population of galaxies in Horizon-AGN being more aligned. Because $\eta_e(r)$ captures alignments around galaxies, this difference could come from an actual difference in alignments in the two simulations, or a difference in how the galaxy population traces the density field (a change of bias, or a more locally anisotropic distribution of galaxies), or both. Similar results are obtained in the projected correlation function at $z=0$ (bottom right panel), where Horizon-AGN shows an excess of radial alignments particularly at small scales.

To distinguish between the two alternative interpretations, the top right panel shows the three-dimensional alignments of the {\it matched} population. The difference between Horizon-AGN and Horizon-noAGN is similar to that seen in the top left panel of Figure \ref{fig:autocorrel_nocut}.
As the matched galaxies live at the same positions in the two simulations, differences in the measured alignment correlation cannot be attributed to the way in which galaxies trace the underlying density field, but to actual changes in their relative orientations, with these being more strongly correlated in the AGN feedback simulation.

The bottom left panel of Figure \ref{fig:autocorrel_nocut} shows projected alignments, i.e. $w_{g+}$, at $z=0$. Similarly to the 3D case, we find that alignments of galaxy shapes around galaxy positions are stronger in projection when implementing AGN feedback. Finally, the bottom right panel shows the $w_{\delta +}$ correlation of all projected galaxy shapes around the matter field. We use this statistic as a proxy for weak lensing contamination.
We find that this correlation is robust to the inclusion of AGN feedback, with similar results in both simulations within our error bars.
This suggests that while the impact of AGN feedback needs to be considered in the modelling of galaxy position-shape correlations ($w_{g+}$), its impact on $w_{\delta +}$, and thus on the contamination of intrinsic alignments to weak lensing, is negligible.

\begin{figure}
    \centering
    \includegraphics[width=\columnwidth]{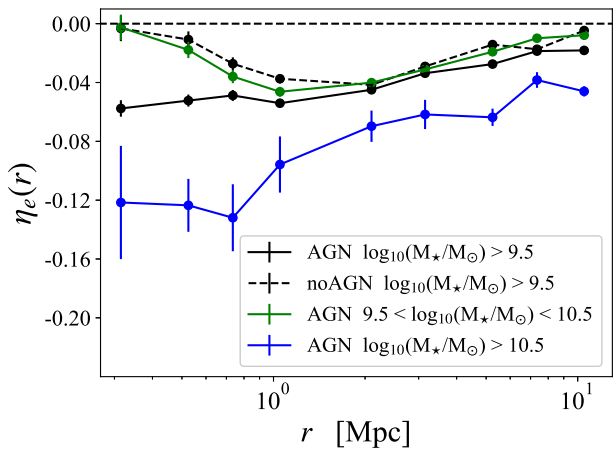}
    \caption{Minor axis-direction of separation correlation for ellipsoids in Horizon-AGN (solid lines) and Horizon-noAGN (dashed line) at $z=0.06$ and according to their stellar mass. Selection criteria for the galaxies include the usual $N>300$ and level$=1$ and ellipsoids are defined as $V/\sigma<0.38$. Note that black lines are the same as the upper left panel of Figure \ref{fig:autocorrel_lowVsigmahighM}}
    \label{fig:autocorrel_lowVsigmaBYMASS}
\end{figure}

\begin{figure*}
    \centering
    \includegraphics[width=7.5cm]{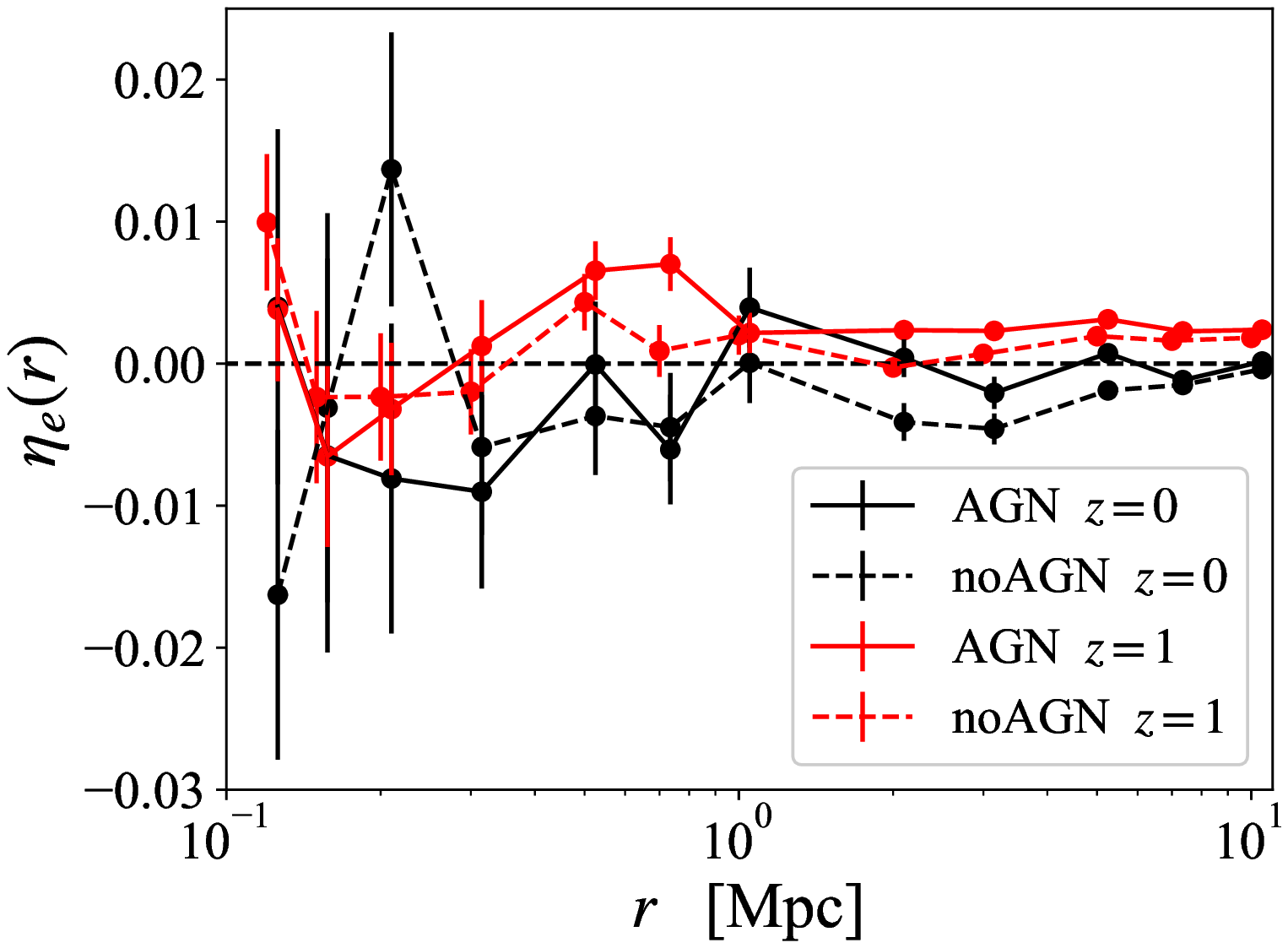}
    \includegraphics[width=7.5cm]{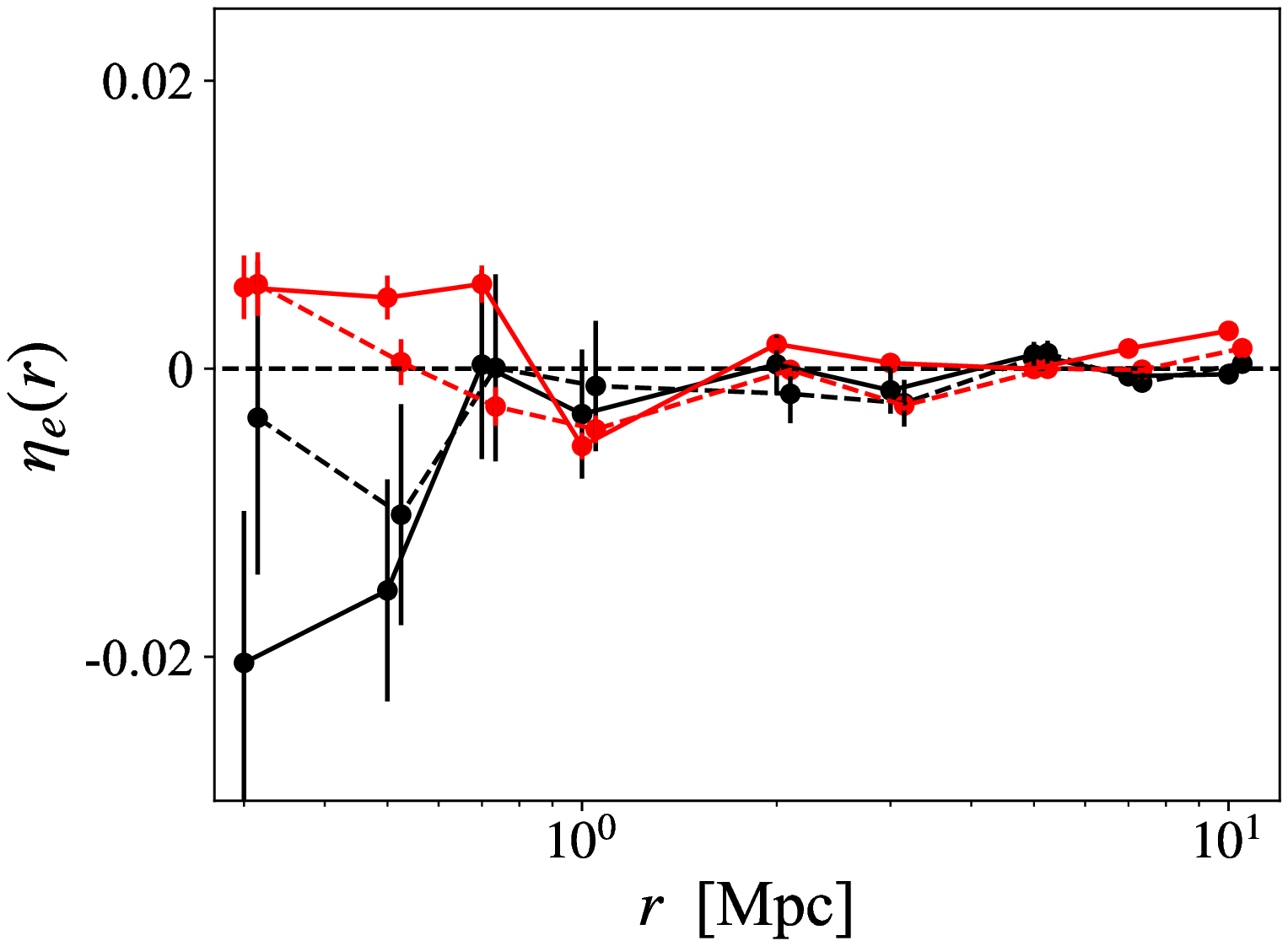}
    \includegraphics[width=7.5cm]{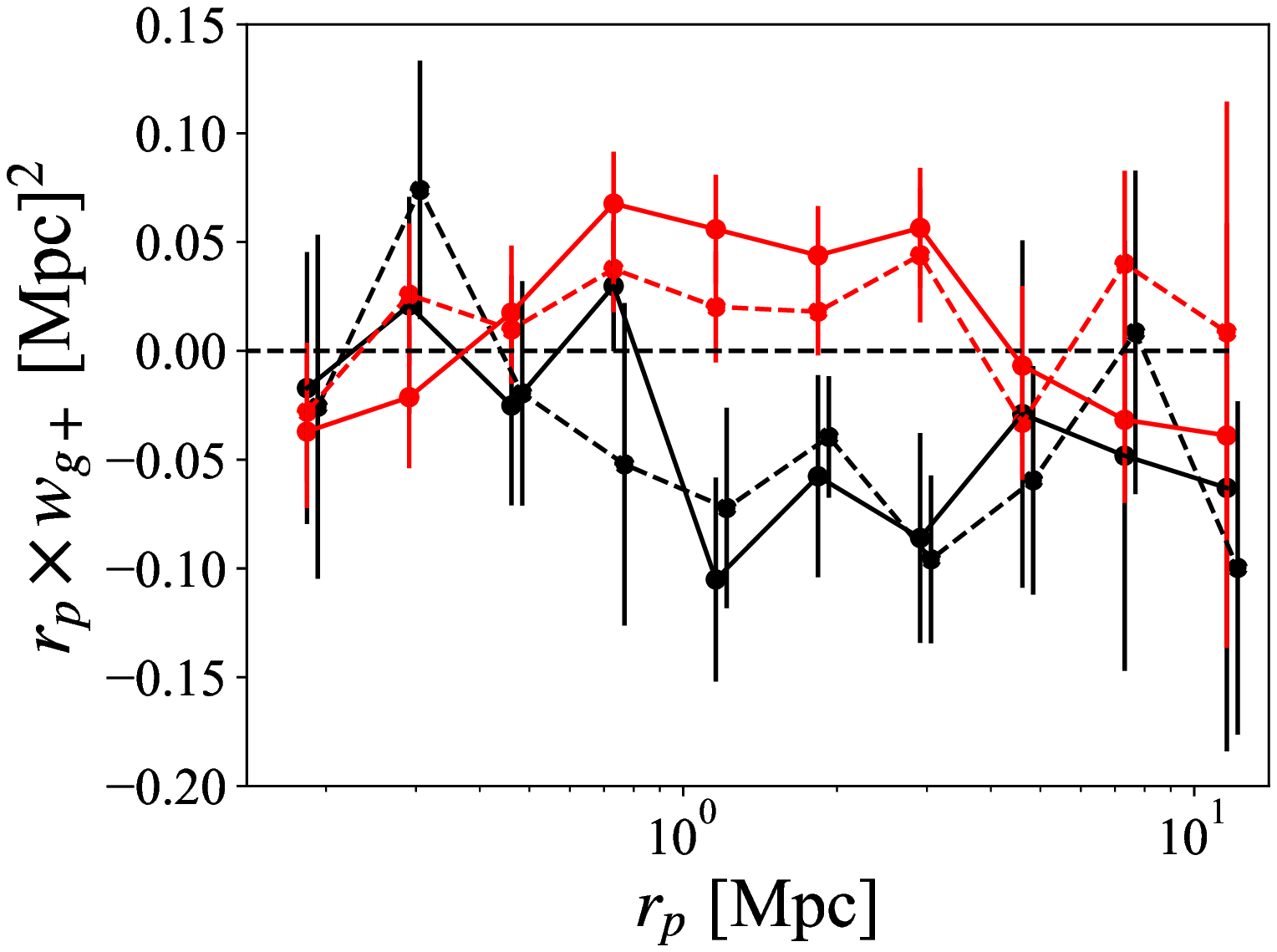}
    \includegraphics[width=7.5cm]{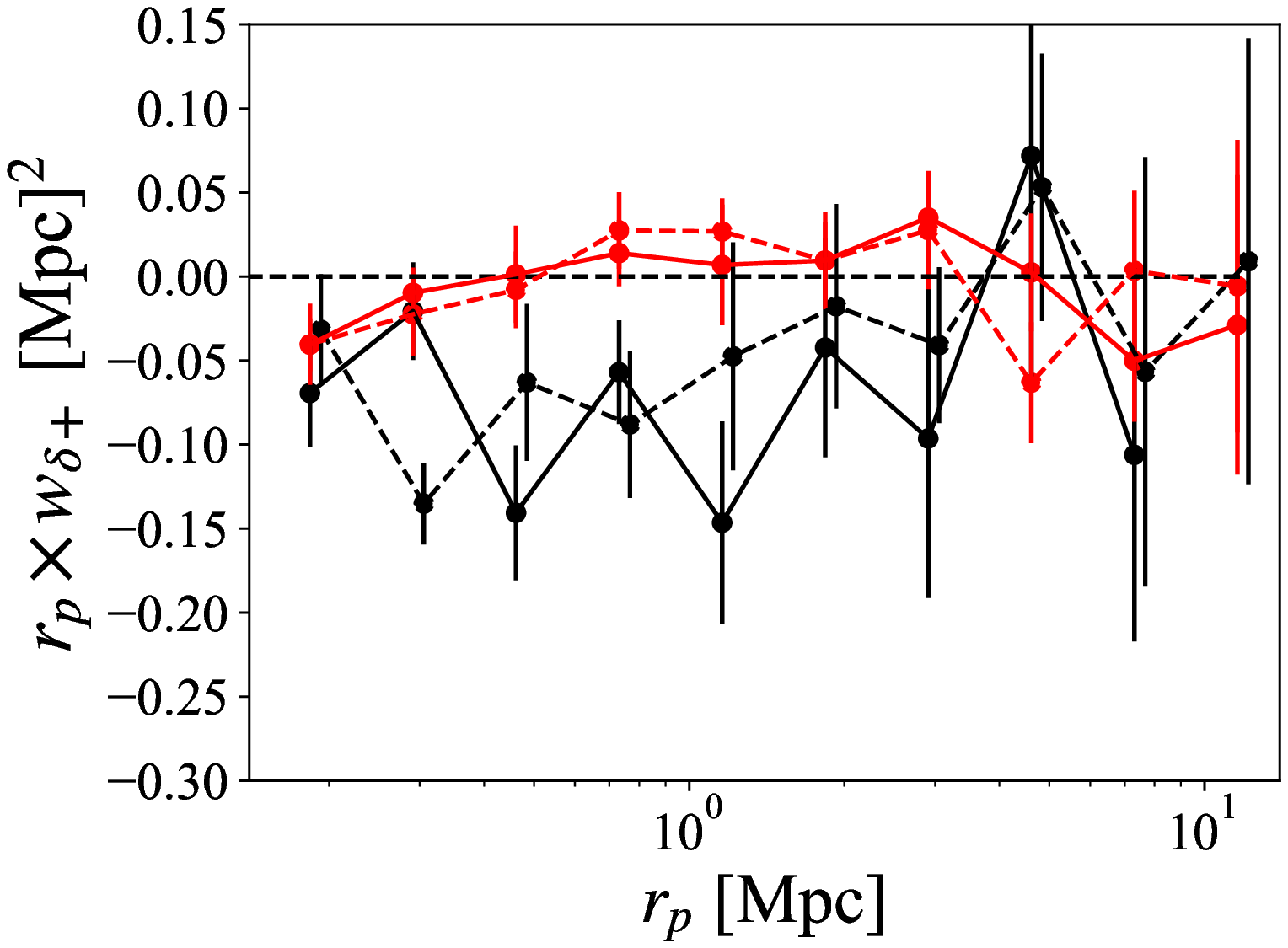}
    \caption{Alignment signals for disc galaxies in Horizon-AGN (solid lines) and Horizon-noAGN (dashed lines). Discs are defined as $V/\sigma>0.6$. Upper left is $\eta_{e}(r)$ at redshift 0 (black lines) and 1 (red lines), upper right is $\eta_{e}(r)$ for the matched galaxy population with cuts made in Horizon-AGN, at z=0 (black lines) and z=1 (red lines). Bottom panels correspond to projected correlations multiplied by the projected correlation distance: $r_{p} \times w_{g+}$ (bottom left) and $r_{p} \times w_{\delta+}$ (bottom right) at $z=0$ (black lines) and $z=1$ (red lines).}
    \label{fig:autocorrel_highVsigma}
\end{figure*}

\subsubsection{High-mass ellipsoids}

Galaxy alignments have only been conclusively observed for massive red pressure-supported galaxies \citep{Mandelbaum06,Hirata07,2011A&A...527A..26J,Li13,Singh14}. These are considered to be the main source of contamination to weak gravitational lensing observables in current surveys. We explore how AGN feedback affects this particular population in Figure \ref{fig:autocorrel_lowVsigmahighM}. This figure is analogous in its layout to Figure \ref{fig:autocorrel_nocut}, except for us selecting only high-mass ellipsoids in both simulations.

The top left panel of Figure \ref{fig:autocorrel_lowVsigmahighM} demonstrates that the minor axes of ellipsoids are oriented perpendicularly to the separation vector. The strength of the alignment is higher than for the whole galaxy population (compare to the top left panel of Figure \ref{fig:autocorrel_nocut}). The redshift evolution however seems to be reversed: high-mass ellipsoids at $z=1$ are rather more aligned than their counterparts at $z=0$. The $w_{g+}$ (bottom left panel) correlations display analogous trends to the ones seen in $\eta_{e}$ at $z=0$. Moreover, the difference between the alignment signals of Horizon-AGN and Horizon-noAGN becomes more prominent in both statistics, compared to Figure \ref{fig:autocorrel_nocut}. The small number statistics at $z=1$ (given our specific selection cut and the redshift evolution of the population - see Table \ref{tab:1}) implies that we cannot reliably identify any difference in the alignments of high-mass ellipsoids between Horizon-AGN and Horizon-noAGN at this redshift.

Compared to the whole galaxy population, high-mass ellipsoids exhibit more important differences in alignments between Horizon-AGN and Horizon-noAGN both in $\eta_{e}(r)$ and $w_{g+}$. This is a result of the presence of very high-mass ($\log_{10}(M_*/{\rm M}_{\odot})>10.5$) ellipsoids in Horizon-AGN that do not exist in Horizon-noAGN. The counterparts in Horizon-noAGN of these very high-mass ellipsoids have higher $V/\sigma$, and therefore are not included in this sample. The Horizon-AGN simulation includes $2143$ ellipsoids $\log_{10}(M_*/{\rm M}_{\odot})>10.5$, while only $245$ of this type exist in Horizon-noAGN. In Figure \ref{fig:autocorrel_lowVsigmaBYMASS}, we reproduce the two curves of Figure \ref{fig:autocorrel_lowVsigmahighM} top left panel at $z=0$ and we divide the sample of high-mass ellipsoids between those of very high mass ($\log_{10}(M_*/{\rm M}_{\odot})>10.5$ and the rest). As shown on Figure \ref{fig:autocorrel_lowVsigmaBYMASS}, those with $\log_{10}(M_*/{\rm M}_{\odot})>10.5$ display a very strong tangential alignment signal of their minor axes, while smaller mass ellipsoids show weaker alignments.

\begin{figure*}
\centering
\includegraphics[width=7.5cm]{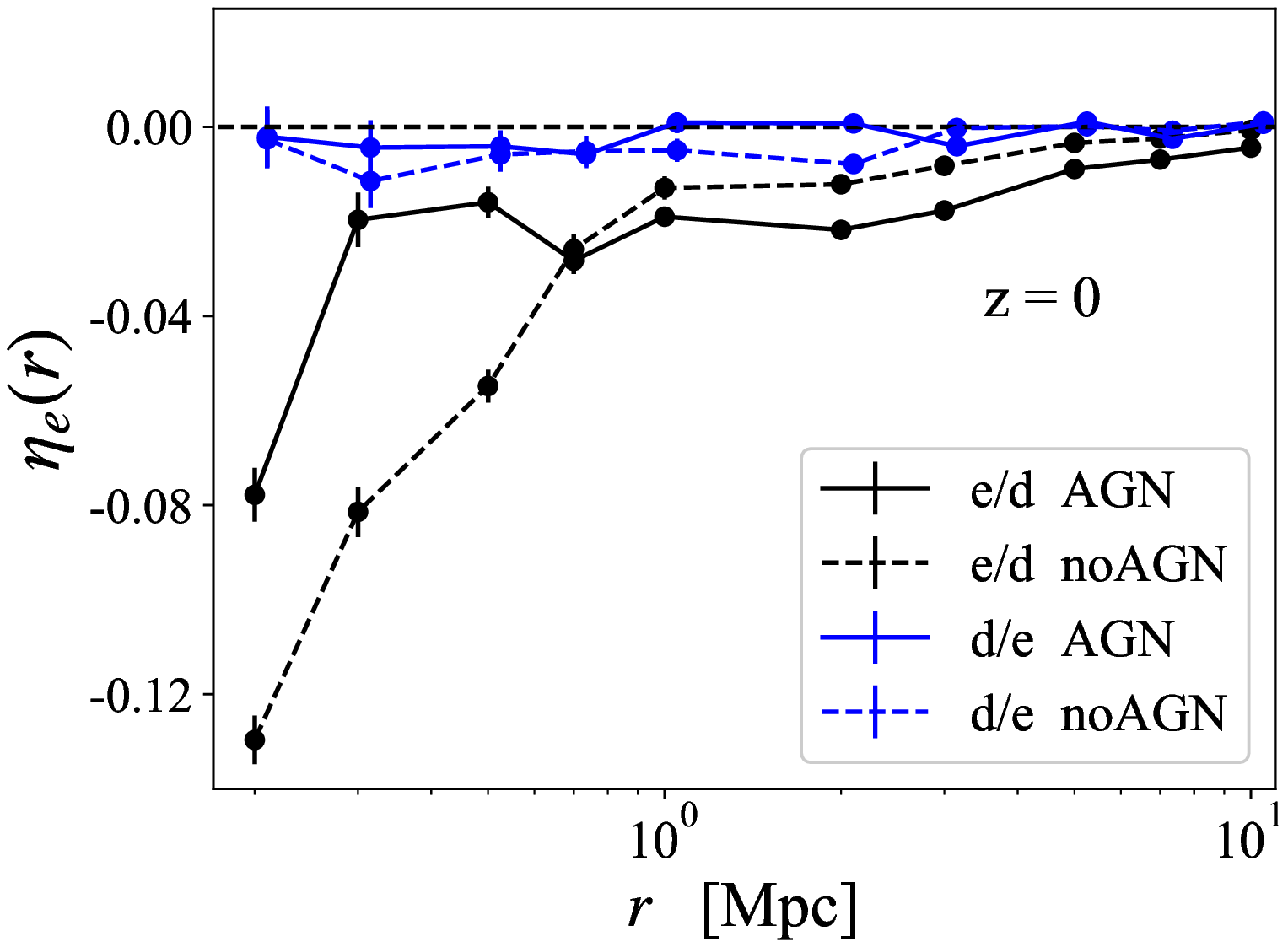}
\includegraphics[width=7.5cm]{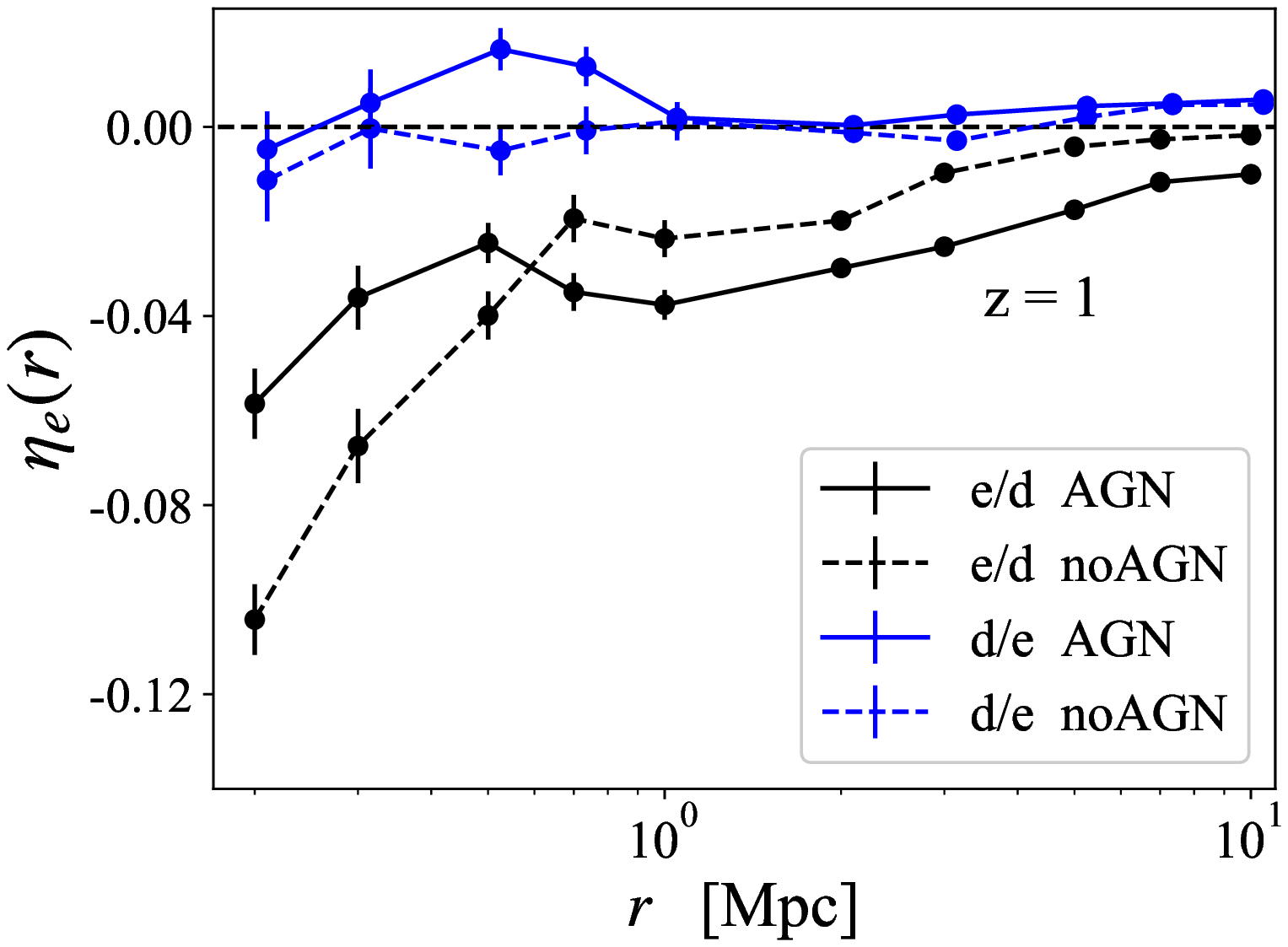}
\includegraphics[width=7.5cm]{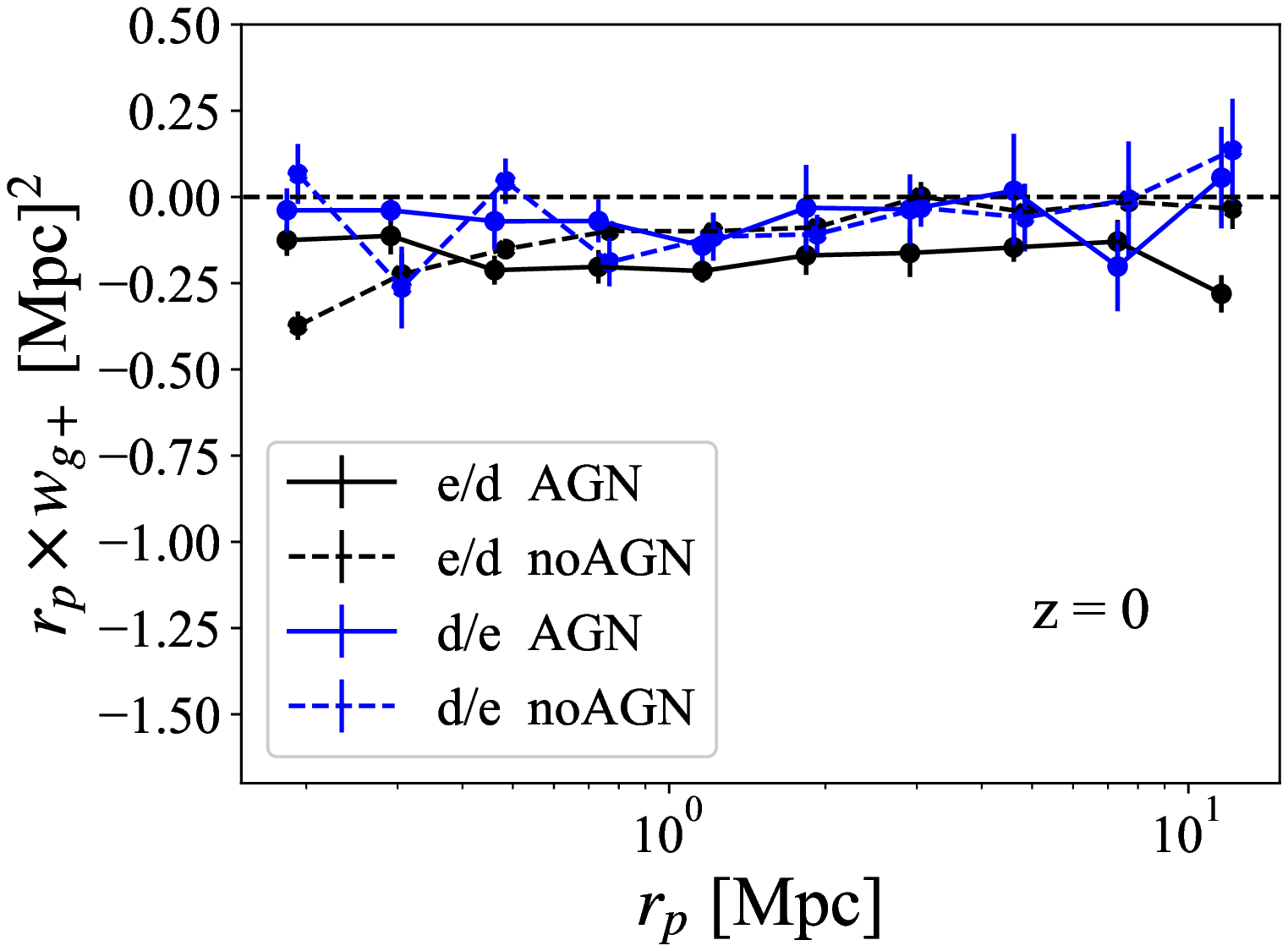}
\includegraphics[width=7.5cm]{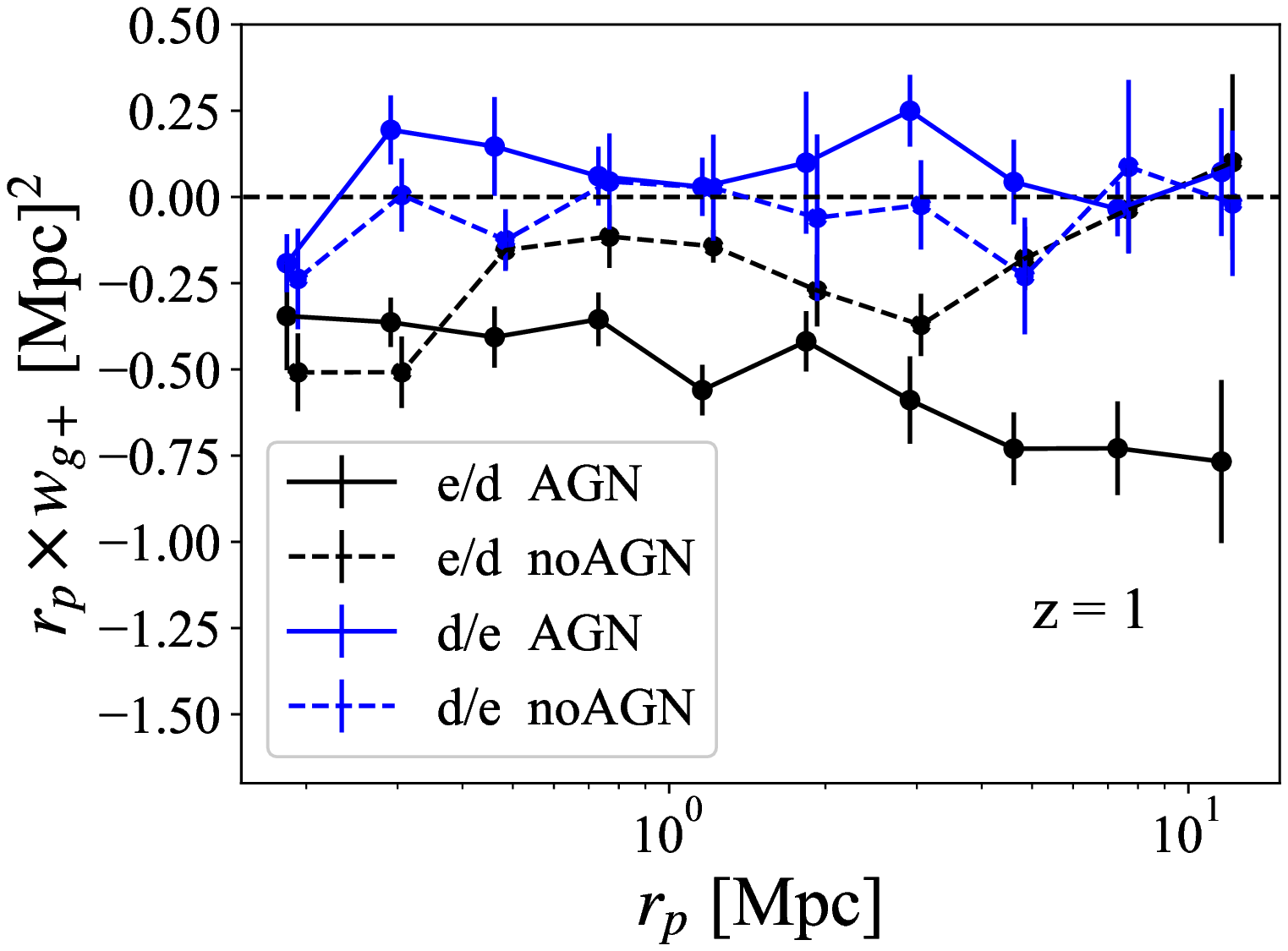}
\caption{Orientation-separation correlation in 3D ($\eta_{e}$, top panels) and in projection ($w_{g+}$, bottom panels) in Horizon-AGN population (solid lines) and Horizon-noAGN (dashed lines) at $z=0$ (left panels) and 1 (right panels). ``e/d'' stands for aligments of ellipticals around discs, and viceversa.}
\label{crosscorrel}
\end{figure*}

Similarly to Figure \ref{fig:autocorrel_nocut}, we can identify a {\it matched} population of high-mass ellipsoids. Their alignments in Horizon-AGN and Horizon-noAGN are shown in the top right panel of Figure \ref{fig:autocorrel_lowVsigmahighM}. Cuts are made in Horizon-AGN and we show the alignments of high mass ellipsoids in Horizon-AGN, and of their twins in Horizon-noAGN.
The overall signal in $\eta_e(r)$, though mostly significant, is smaller than for the full population of high-mass ellipticals. The difference between Horizon-AGN and Horizon-noAGN is also less striking. These small signals could be a consequence of the matching: as shown on Figure \ref{fig:matchingfractions}, matching fractions are low at low $V/\sigma$, hence it is possible that the high-mass ellipsoids responsible for the signal have not been successfully matched ($50\%$), accounting for a decrease in the alignment signal.

As for $w_{\delta +}$, shown in the bottom right panel of Figure \ref{fig:autocorrel_lowVsigmahighM}, the alignment signal when considering all massive ellipsoids is again consistently stronger than when considering the whole galaxy population. Some discrepancies can be seen at large scales, as opposed to the case of the full population in Figure \ref{fig:autocorrel_nocut}.

\subsubsection{Discs}

Our previous studies with Horizon-AGN have suggested disc galaxies are also subject to alignments, but with a small trend to align their minor axes radially around over-densities in the matter field. Other simulations have predicted the opposite behavior and the source of discrepancy has not yet been elucidated \citep{Tenneti16}. It is thus of relevance to explore whether the presence of AGN, known to change the morphology and internal dynamics of galaxies, can be a factor driving these results.

In Figure \ref{fig:autocorrel_highVsigma}, we show our results for disc ($V/\sigma>0.6$) shape alignments. Alignments both in three-dimensions and projected are found to be mostly consistent with null, though $w_{\delta+}$ and $w_{g+}$ exhibit slightly tangential trends. In comparison, \citet{2015MNRAS.454.2736C} had found a radial trend for the auto-correlations of discs using the reduced inertia tensor rather than the simple one adopted in this work. The use of the simple inertia tensor to trace disc shapes results in slightly tangential signals, qualitatively consistent with recent observational results \citep{Georgiou19}. We touch upon the impact of the choice of inertia tensor for tracing the shape of discs in Appendix \ref{inertiachoice}, while correlations of their spins are produced in Section \ref{sec:spin}.

Alignments of discs in Horizon-AGN and Horizon-noAGN are typically consistent within error bars for $\eta_{e}(r)$, $w_{g+}(r_{p})$ and $w_{\delta +}(r_{p})$. Hence, AGN feedback is not responsible for damping any preexisting alignments: alignments of discs around themselves are already close to zero without AGN feedback.

\subsection{Cross-correlations} \label{Crosscorrelations}

We also explore the influence of AGN feedback on cross-correlations between different samples  of galaxies. Such correlations are beginning to be used in observational analyses of intrinsic alignments to constrain the modeling \citep{Samuroff18}. Moreover, our previous work with Horizon-AGN showed that the alignment signal of discs was more prominent when computed around more biased tracers of the density field, such as ellipsoids. Hence, it is more likely that we can see an impact of AGN on disc alignments in cross-correlation with ellipsoids. In this section, we use all ellipsoids with reliable shape measurements in the simulation, that is, all galaxies with $V/\sigma<0.38$.

Figure \ref{crosscorrel} depicts cross-correlations between discs and ellipsoids at $z=0$ and $z=1$. Alignments of discs around ellipsoids at $z=0$ (``d/e'', red curves in every panel) do not present significant differences between Horizon-AGN and Horizon-noAGN, and are mostly consistent with $0$ when using the simple inertia tensor.

Alignments of ellipsoids around discs (``e/d'') are radial in Horizon-AGN at $z=0.5$ \citep{2015MNRAS.454.2736C}. Here, we verify this trend at $z=0$ and $z=1$. These three-dimensional alignments correspond to the strongest signal measured so far in any of the previous figures. (We can however observe even stronger alignments when including secondary structures, as discussed in Appendix \ref{app:levels}.) Interestingly, at both redshifts, the implementation of AGN feedback damps the alignments at small scales ($r<0.8$ Mpc) by a factor of up to 3, while it amplifies them at large scales ($r>0.8$ Mpc) by a factor of up to 2. Once again, this pattern is reproduced in projection, with Horizon-noAGN signal being dominant at low correlation length for both redshifts, while Horizon-AGN dominates at higher correlation length.

\begin{figure}
\centering
\includegraphics[width=\columnwidth]{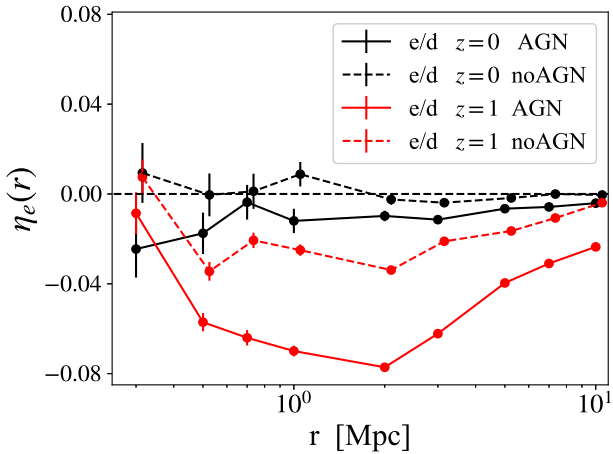}
\caption{Cross-correlation of ellipsoids ($V/\sigma<0.38$) around discs ($V/\sigma>0.6$) (e/d) in the matched galaxy population at $z=0$ (black lines) and $z=1$ (red lines) in Horizon-AGN (solid lines) and Horizon-noAGN (dashed lines).}\label{fig:matchedcross}
\end{figure}

The differences in alignment trends described above could be due to a difference in the galaxy distribution (a more anisotropic distribution of galaxy positions for instance), as well as a modification of the way ellipsoids align around discs when AGN feedback is present in the simulation. To distinguish between these two options, we once again turn to the matched sample, whose alignments are shown in Figure \ref{fig:matchedcross}. Interestingly, in the matched population, and contrarily to the non-matched one, the Horizon-AGN alignment signal dominates over the Horizon-noAGN one at any correlation length, particularly at $z=1$. The persistence of a difference in the alignment signal when using these matched samples highlights the fact that AGN feedback directly modifies the way ellipsoids align around discs. At small scales, the fact that using non-matched samples the Horizon-noAGN signal dominates while using matched samples the Horizon-AGN signal is dominant reveals that the small-scale alignment difference observed in the matched sample is due to a difference in the galaxy bias across simulations. More precisely the small scale distribution of ellipsoids around discs must be more anisotropic in Horizon-noAGN than in Horizon-AGN.

\begin{figure*}
\centering
\includegraphics[width=7.5cm]{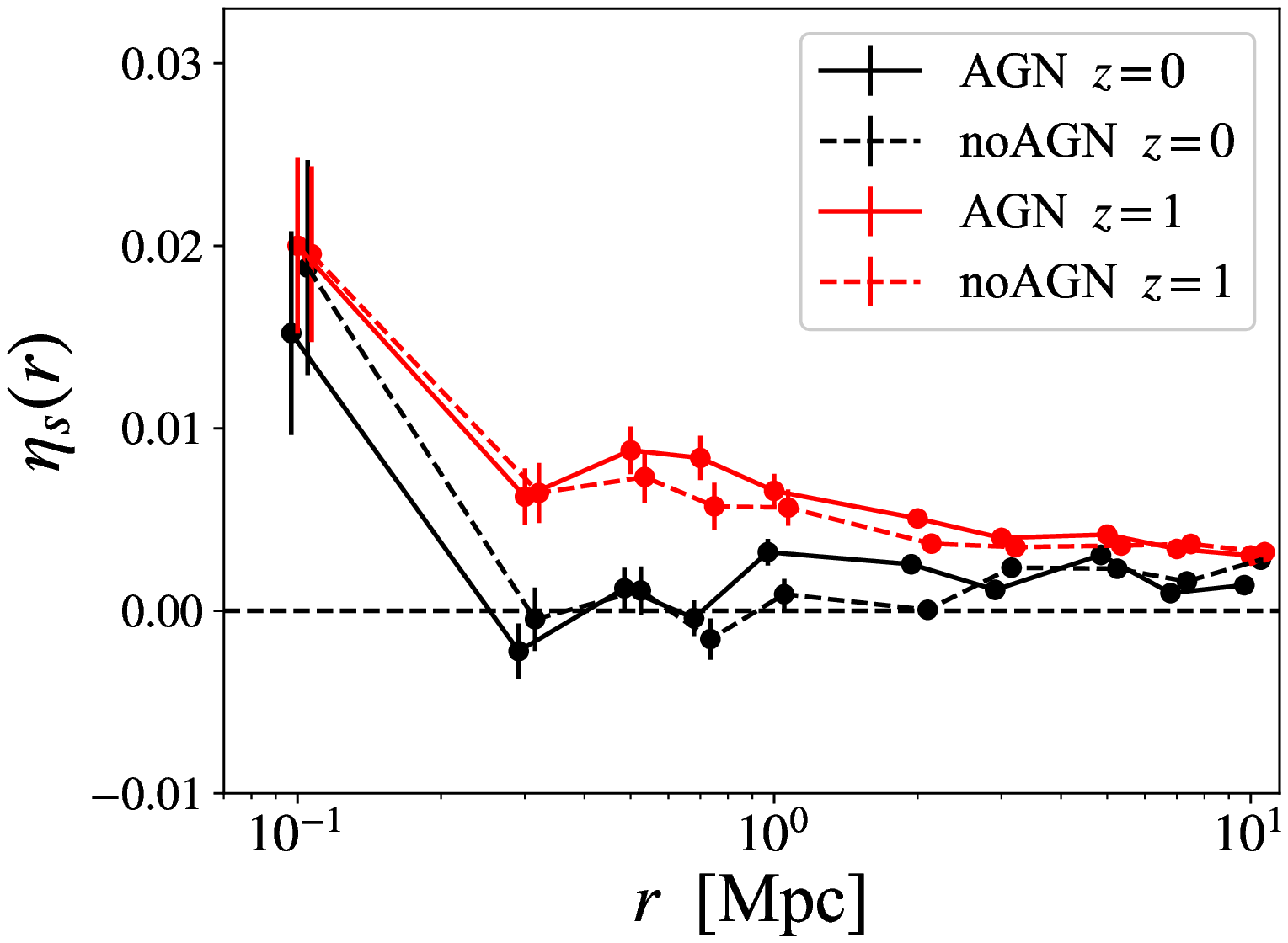}
\includegraphics[width=7.5cm]{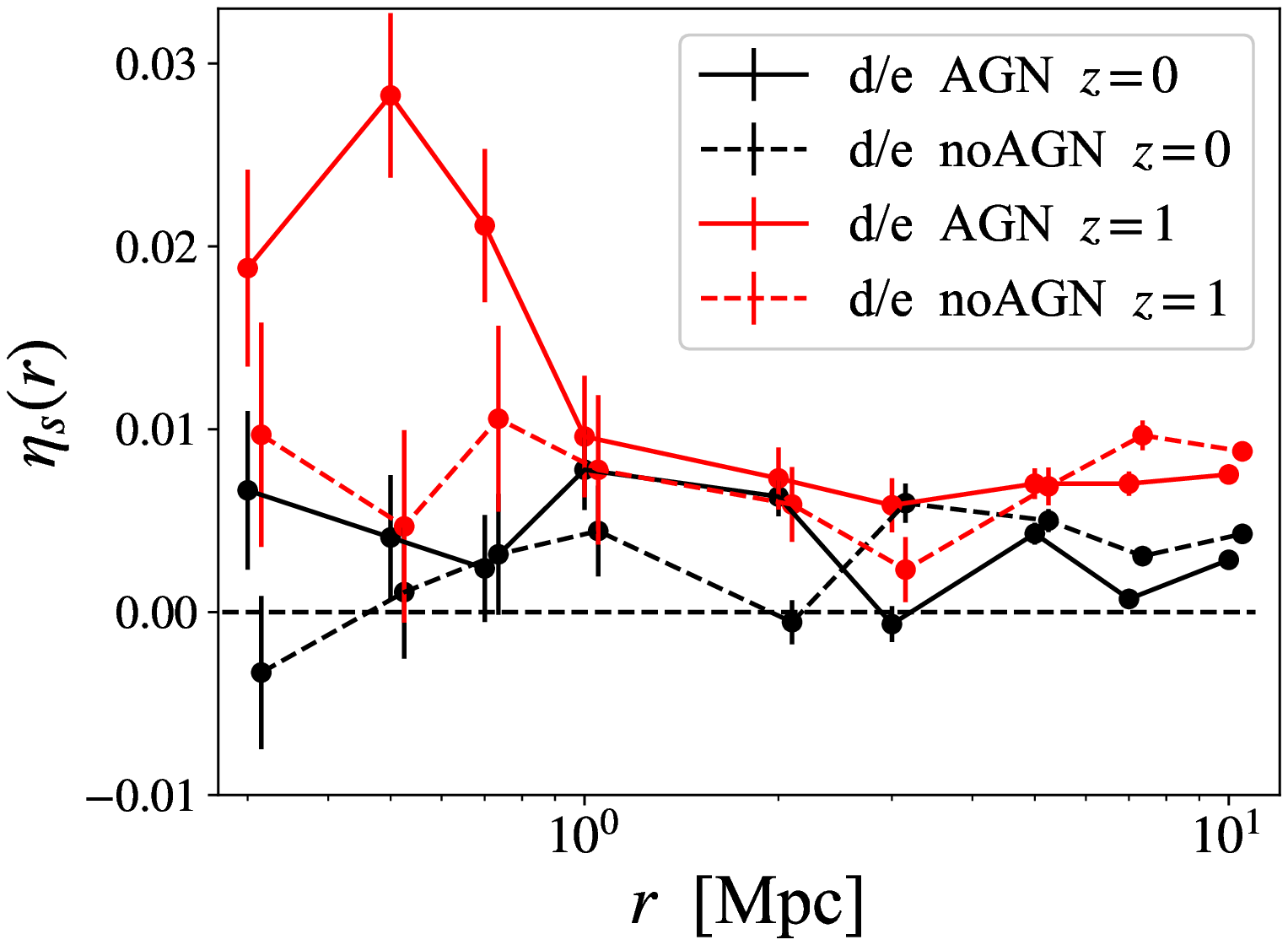}
\caption{Spin-separation auto-correlation of the whole galaxy population (left) and spin-separation cross-correlation of discs ($V/\sigma>0.6$) around ellipsoids ($V/\sigma<0.38$) in Horizon-AGN (solid lines) and Horizon-noAGN (dashed lines) at $z=0$ (black lines) and $z=1$ (red lines).}
\label{fig:spincorrel}
\end{figure*}

\subsection{Spin correlations}
\label{sec:spin}

So far we have only considered galaxy shape alignments. However galaxies can suffer tidal torquing, which creates alignments of their spin \citep{10.1046/j.1365-8711.2001.04105.x,Hui2002, ATTT}.
Moreover alignments of spins around other galaxies and with the cosmic web have been tentatively detected in observations \citep{2000ApJ...543L.107P, 10.1111/j.1365-2966.2008.13655.x, 10.1111/j.1365-2966.2010.17202.x, 10.1111/j.1365-2966.2011.19620.x, Tempel:2013gqa} and correlations of spins of galaxies have been found to be non-zero in Horizon-AGN \citep{codis14,2015MNRAS.454.2736C}. In this section, we assess the role played by AGN feedback in spin alignments in this simulation.

In Figure \ref{fig:spincorrel}, we present auto-correlations of spins in the whole galaxy sample (left panel) and cross-correlations of the spins of discs around ellipsoids (right panel). The left panel shows that the auto-correlation signal of the full galaxy population is radial and higher at $z=1$ (as expected from \citealt{2015MNRAS.454.2736C}). Moreover, the signals from Horizon-AGN and Horizon-noAGN are consistent within error bars at both redshifts and for most of the $r$ range. The same is true for the correlation of the spins of discs around ellipsoids are tangential.

While at $z=0$ Horizon-AGN and Horizon-noAGN exhibit consistent spin alignment signals, at $z=1$ there is a small indication of higher alignments in Horizon-AGN, though the limited statistics of the box prevent us from drawing a firm conclusion. The spin alignments of discs around ellipsoids at $z=0$ fairly reproduce the trends found when using the reduced inertia tensor minor axis in Appendix \ref{inertiachoice}, highlighting the fact that the reduced inertia tensor minor axis can serve as tracer of disc spins, as shown in \citet{2015MNRAS.454.2736C} and \citet{2017MNRAS.472.1163C}.

\subsection{Alignments with filaments}
\label{sec:filamentsIA}

\citet{ATTT} showed that the net effect of the cosmic web was to set preferred directions for the orientation of spins, with more massive galaxies having a spin perpendicular to the nearest filament while less massive galaxies have a spin parallel to it. This spin flip has been measured for spins of haloes in dark-matter only simulations \citep{BailinSteinmetz,Aragon_Calvo_2007,Codis2012,Trowland2013} and was interpreted as a result of mergers during the flow towards the nodes of the cosmic web \citep{Codis2012,Welker14,WangKang17}. To account for baryonic physics, \citet{Dubois14} and then \citet{Codis2018} investigated spin orientations with respect to filaments in the Horizon-AGN simulation, and found a similar spin flip among the galaxies of Horizon-AGN. \citet{Dubois14} argued that AGN feedback played a key role in preventing parallel alignment from being re-gained by cold gas re-accretion after a merger. AGN feedback would ensure that the galaxy alignment is dominated by its ex-situ stellar component. We here seek to further isolate the role of AGN feedback in the alignment of spins with filaments by comparing these alignments in the Horizon-AGN and Horizon-noAGN simulations.

We compute the probability distribution function of the cosine of the angle between the spin of individual galaxies and the direction of the nearest filament in different stellar mass bins. We study all galaxies in Horizon-AGN binned by mass, all galaxies in Horizon-noAGN binned by mass, and matched galaxies in Horizon-noAGN binned by mass of their twins in Horizon-AGN.

\begin{figure*}

\centering
\includegraphics[width=18cm]{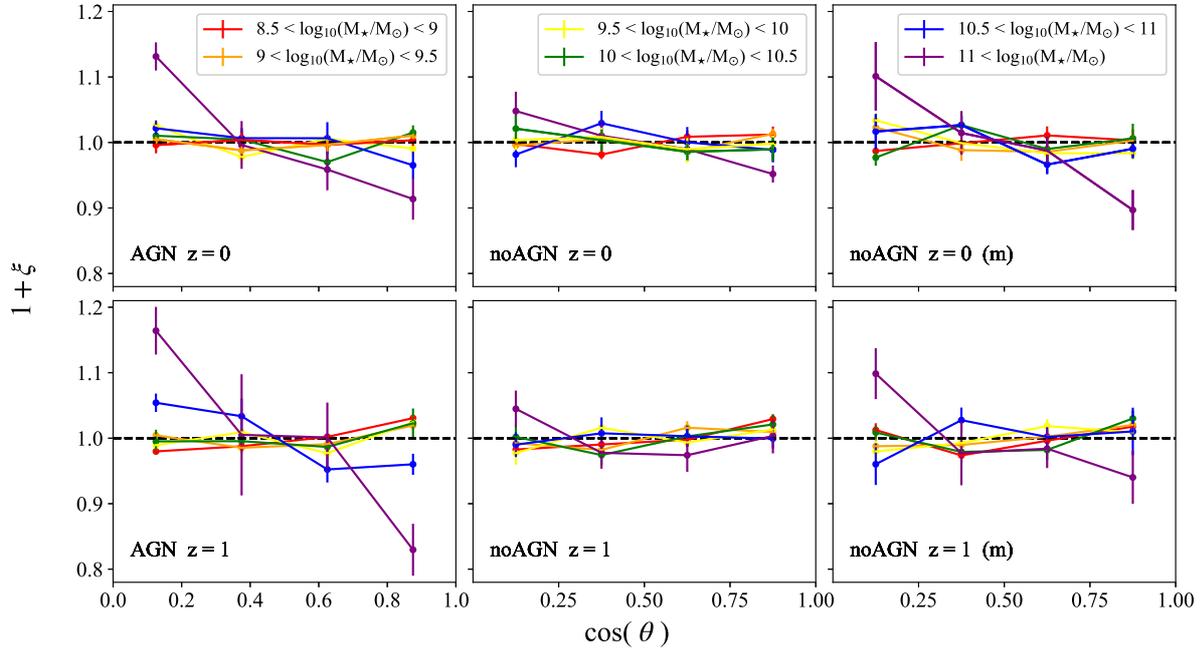}
\caption{Probability distribution function of the cosine of the angle between spins of galaxies and the direction of the nearest filament for 6 stellar mass bins in Horizon-AGN (left column), Horizon-noAGN (middle column) and twins of Horizon-AGN galaxies in Horizon-noAGN (right column, (m) stands for matched), at $z=0$ (top row) and $z=1$ (bottom row). In the right column, we study the twins in Horizon-noAGN of galaxies within each mass range in Horizon-AGN.} \label{fig:filaments}
\end{figure*}

As depicted on Figure \ref{fig:filaments}, we observe the same spin alignment transition as \citet{Codis2018} in the Horizon-AGN population. The most massive galaxies at $z=0$ and $z=1$ ($\log_{10}(M_{*}/{\rm M}_{\odot})>11$, purple curves) have an excess of probability to have low values of $\cos(\theta)$, that is, to be at an angle of $\pi/2$ with respect to their nearest filament. This excess of alignment is also detected to a lower extent in the $10.5<\log_{10}(M_{*}/{\rm M}_{\odot})<11$ mass bin. On the other hand, low mass galaxies ($8.5<\log_{10}(M_{*}/{\rm M}_{\odot})<9.5$, red curves) have a tendency to be aligned parallel to the direction of the nearest filament with an excess of probability for $\theta$ close to 0 only at $z=1$.

This spin transition observed in Horizon-AGN is less significant in Horizon-noAGN, as showcased in the middle panels of Figure \ref{fig:filaments}. At both redshifts, the most massive galaxies only show a slight excess of probability to be perpendicular with respect to the filament, while the least massive galaxies display an excess of probability to be aligned comparable to the one found in Horizon-AGN. This demonstrates that AGN feedback has a more dramatic influence on higher mass galaxies.

The right panels of Figure \ref{fig:filaments} focus on the matched sample. Twins of massive galaxies display similar signals in Horizon-AGN and noAGN, albeit the significance is lower for galaxies with masses between $10^{10.5}$ and $10^{11}$ M$_{\odot}$. On the other hand, twins of low mass galaxies do not exhibit any significant trend towards alignment with filaments. Looking in more detail at all the twins in Horizon-noAGN of galaxies with $\log_{10}(M_{*}/{\rm M}_{\odot})>11$ in Horizon-AGN, we find that these also have masses above that threshold in Horizon-noAGN. This suggests that, in the middle panel, the high mass bin corresponds to the galaxies of the right panel plus another component in Horizon-noAGN which are either not matched or matched to galaxies with lower mass in Horizon-AGN. Those galaxies exhibit an alignment signal with the filament that is consistent with null but are rather numerous ($\sim$ 3500 compared to only $\sim$ 2000 twins). Their presence decreases the significance of the perpendicular alignment signal of high-mass galaxies.

Our results support the conclusions of \citet{Dubois14} on the role of AGN feedback on spin alignments with filaments. In its absence, galaxies grow to higher masses by accreting gas in manner that allows them to preserve the parallel spin alignment with filaments.

\section{Conclusions}
\label{sec:Conclusions}

In this work, we have isolated the impact of AGN feedback on galaxy alignment using two simulations of the Horizon suite. The only difference between the two simulations is the implementation of an AGN feedback (with thermal and kinetic modes) in Horizon-AGN, which is missing in Horizon-noAGN. Our main findings are as follows.

\begin{itemize}
    \item AGN feedback does not change the alignments of the overall population of galaxies around the matter density field in the simulations. Hence, within our error bars, it does not impact the modelling of the main source of alignment contamination in cosmic shear studies, the ``GI" term \citep{2004PhRvD..70f3526H}. In the future, this will need to be confirmed by making a more representative selection of galaxies in the simulation specifically designed to match that of lensing studies with current and future surveys.
    \item AGN feedback changes the correlation of galaxy shapes around biased tracers of the density field (i.e., galaxy positions). As this is the main observable used to measure intrinsic alignments directly and constrain their modelling, our results suggest that this effect needs to be taken into account for such applications and particularly at small scales. At such scales, other elements of sub-grid modelling in hydrodynamical simulations might also play a role and should be investigated in existing suites.
    \item Galaxy matching across simulations allowed us to isolate the effect of AGN on galaxy orientations from selection effects driven by our choice of galaxy samples (ie., the fact that galaxies in Horizon-noAGN are more disc-like). Our results suggest that AGN has an impact on driving galaxy alignment trends through both a change in the distribution of galaxy properties and in the correlation of their orientations. From an observational point of view, this suggests that combinations of such measurements could establish useful constraints on the impact of this feedback mechanism.
    \item  In the comparison done in the matched sample, wherever differences of alignments around biased tracers of the density field are observed, alignment signals are reinforced when implementing AGN feedback. This is a consequence of the presence of more pressure-supported galaxies, known to be the main population subject to intrinsic alignments.
    \item Within our error bars, spin alignments of the overall galaxy population are not modified by AGN feedback. Neither are the spin alignments of matched samples. The only changes perceived in spin alignments are in the cross-correlations of non-matched samples, suggesting they are due to changes in the population across selection boundaries rather than actual changes in their alignment.
    \item Spin alignments around filaments are weaker when considering high-mass galaxies in Horizon-noAGN than when considering similar mass galaxies in Horizon-AGN. However, this damping is due to a difference in the distribution and abundance of high-mass galaxies, as signals are found to be consistent when considering the matched sample.
\end{itemize}

All in all, our main findings are that, within our error bars, AGN feedback does not impact the matter-shape correlations that are used to estimate contamination to weak gravitational lensing surveys. This should be verified in the future by better mimicking the selection of galaxies in a typical weak lensing survey. However, AGN feedback impacts the position-shape correlations, often used as the relevant observable to constrain intrinsic alignment models. It remains to be seen whether this effect is currently distinguishable from observations. Propagating our findings to forecasts is the next natural step.

While the Horizon-AGN simulation is successful in reproducing observed luminosity functions, mass functions, colours and morphologies of galaxies \citep{Dubois16,Kaviraj17}, its AGN feedback is not sufficient to bring the fraction of gas in haloes to agreement with observed values \citep{Chisari18}. Enhancing the efficiency of AGN feedback could impact alignment statistics, although it is difficult to be fully conclusive, given potential degeneracies between sub-grid parameters.

The sensitivity of intrinsic alignments to AGN feedback suggests they could become a probe of this effect in the future, either via two-point observables or the analysis of the cosmic web (i.e., the mass transition of alignments with respect to filaments). Conversely, it suggests that care must be taken when using hydrodynamical simulations when interpreting observations of galaxy alignments. Our work highlights the need to explore the parameter space of sub-grid models in hydrodynamical simulations when comparing predictions to observations (see \citealt{Chisari19} for a similar proposal for exploring the impact of baryonic physics on predictions for the cosmic shear). Existing simulation suites where multiple calibration runs are available would already provide valuable tools to understand the interplay of intrinsic alignments and sub-grid physics. This seems particularly timely in the context of multiple weak lensing surveys coming online in the next decade, which will improve constraints on intrinsic alignments and advance the knowledge of the role of AGN feedback on galaxy evolution.

\section*{Acknowledgements}

NEC has been supported by a Royal Astronomical Society Research Fellowship. CL is supported by a Beecroft fellowship. RSB acknowledges funding from the Centre National de la Recherche Scientifique (CNRS) on grant ANR-16-CE31-0011. This work has made use of the HPC resources of CINES (Jade and Occigen supercomputer) under the time allocations 2013047012, 2014047012 and 2015047012 made by GENCI. This work is partially supported by the Spin(e) grants {ANR-13-BS05-0005} (\url{http://cosmicorigin.org}) of the French {\sl Agence Nationale de la Recherche} and by the ILP LABEX (under reference ANR-10-LABX-63 and ANR-11-IDEX-0004-02). We thank S. Rouberol for running  smoothly the {\tt Horizon} cluster for us. Part of the analysis of the simulation was performed on the DiRAC facility jointly funded by STFC, BIS and the University of Oxford.


\bibliographystyle{mnras}
\bibliography{noagnbib}


\appendix

\section{Choice of the inertia tensor}
\label{inertiachoice}

\begin{figure}
\centering
\includegraphics[width=0.45\textwidth]{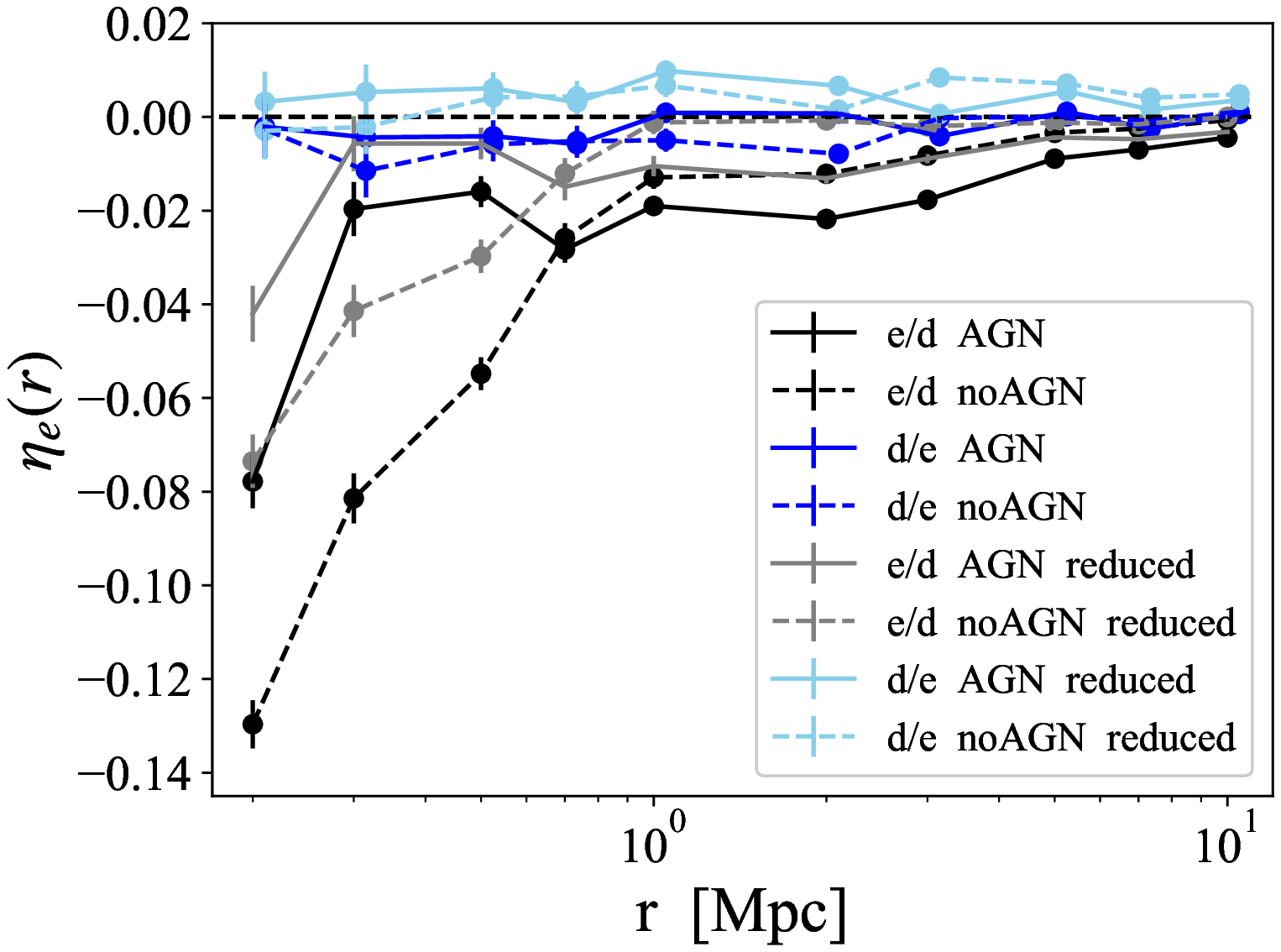}
\caption{Orientation-separation cross-correlations for galaxies with $N>300$ and level=1 using the simple and reduced inertia tensor for Horizon-AGN population (solid lines) and Horizon-noAGN (dashed lines). e/d designates cross-correlations of ellipsoids around discs and d/e of discs around ellipsoids. Ellipsoids correspond to $V/\sigma<0.38$ and discs to $V/\sigma>0.6$.}
\label{simplered}
\end{figure}

When computing alignments, the use of the reduced inertia tensor (Eq. \ref{eq:rit}) tends to lower tangential alignments and to increase radial ones. For instance, Figure \ref{simplered} compares the impact of choice of the inertia tensor in the alignment cross-correlations shown in Figure \ref{crosscorrel}. In Figure \ref{simplered}, the alignments of discs around ellipsoids go from consistent with null to a radial yet weak alignment when using the reduced inertia tensor. On the contrary, tangential alignments of ellipsoids around discs are dampened when using the reduced inertia tensor, by a factor of almost 2 in the most extreme case. However, the choice of tensor does not influence which signal dominates between Horizon-AGN and Horizon-noAGN, and hence our qualitative conclusions when comparing Horizon-AGN and Horizon-noAGN are not affected when using either the simple or reduced inertia tensor.

\begin{figure}
\centering
\includegraphics[width=0.45\textwidth]{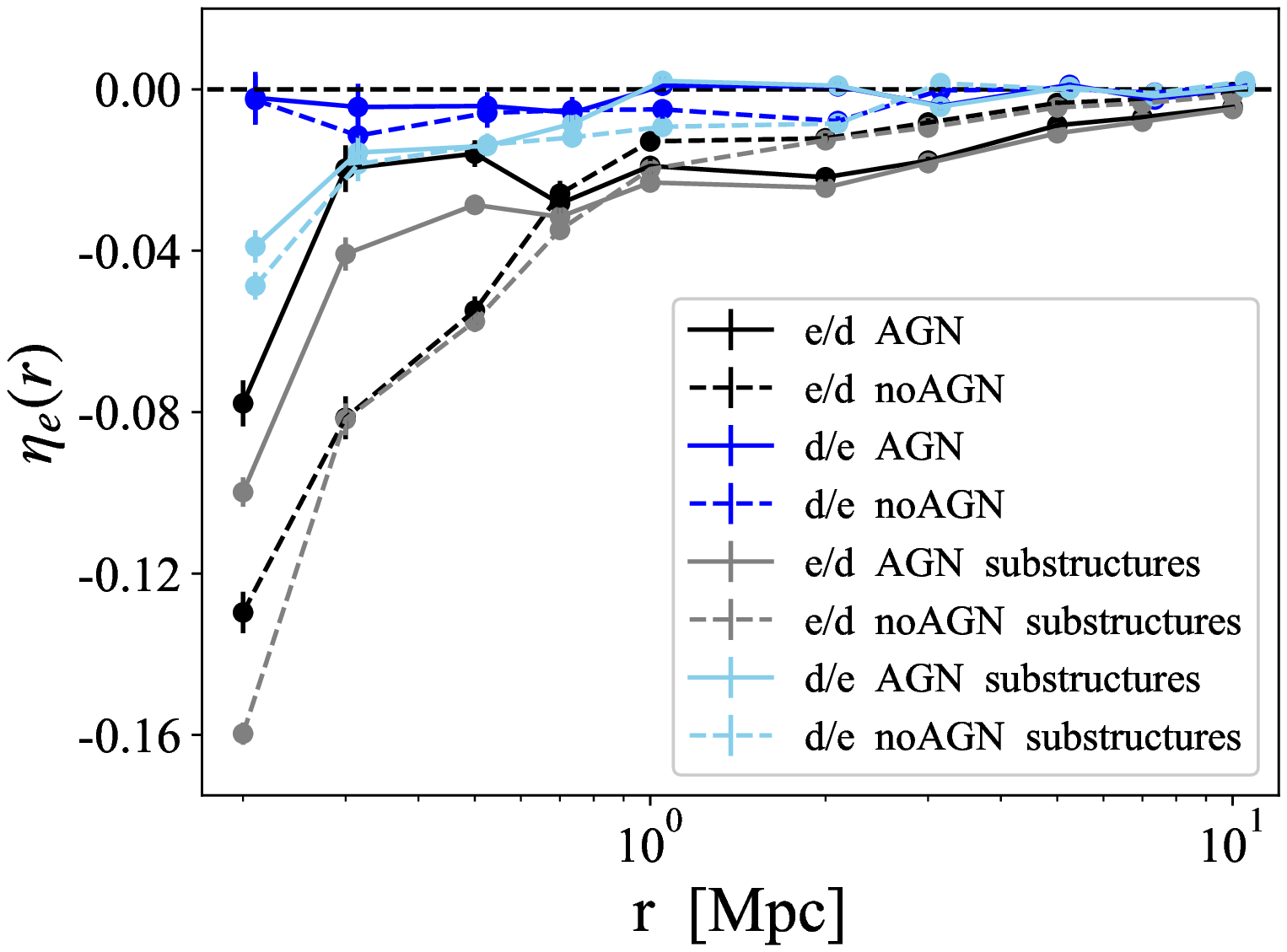}
\caption{Orientation-separation cross-correlations for galaxies with $N>300$ and level=1 or all levels (``substructures"), using reduced inertia tensor. Horizon-AGN population corresponds to solid lines and Horizon-noAGN corresponds to dashed lines). e/d stands for ellipsoids around discs and d/e for discs around ellipsoids. Discs correspond to $V/\sigma>0.6$ and ellipsoids, to $V/\sigma<0.38$.}
\label{levelscross}
\end{figure}

\section{Impact of substructure}
\label{app:levels}
As detailed in Section \ref{HorizonSimulation}, {\sc AdaptaHOP} classifies stellar structures into principal structures (galaxies) to which we assign level$=1$ and secondary and tertiary ones, to which we assign level$=2,3$, respectively. Including secondary structures in alignment statistics influences results significantly at small scales \citep{2016MNRAS.461.2702C}.

Figure \ref{levelscross} compares cross-correlations using only level$=1$ and galaxies of any level. Including secondary structures amplifies negative correlations and turns positive ones into negative. This pattern highlights the fact that secondary and tertiary structures align easily with the primary structure in which they are located, and that these alignments are tangential.


\bsp	
\label{lastpage}
\end{document}